\documentclass[12pt]{iopart}

\usepackage{harvard}
\usepackage{color}
\usepackage[pdftex]{graphicx}
\newcommand{\pd}[2]{\frac{\partial #1}{\partial #2}}
\newcommand{\pda}[2]{\frac{\partial #1 ^\ast}{\partial #2 ^\ast}}

\newcommand{\Rew}{\textrm{Re}_{\rm w}}
\newcommand{\Ret}{\textrm{Re}_{\tau}}
\newcommand{\Wew}{\textrm{Wi}_{\rm w}}

\begin{document}

\title[]{Viscoelastic effect on steady wavy roll cells in wall-bounded shear flow}

\author{
Tomohiro Nimura,
Takuya Kawata, and 
Takahiro Tsukahara\footnote{Corresponding author: tsuka@rs.tus.ac.jp}
}

\address{
Department of Mechanical Engineering, Tokyo University of Science, Chiba, 278-8510, JAPAN
}
\ead{tsuka@rs.tus.ac.jp}

\begin{abstract}
An alternative step in understanding the flows of near-wall drag-reducing turbulence can be examining the flow in a well-organized streamwise vortex with a laminar background.
Herein, we studied the flow behaviors of the Giesekus viscoelastic fluid in a rotating plane Couette flow system, which is accompanied by steady roll-cell structures.
By fixing the Reynolds and rotation numbers at values that provide as steady wavy roll cell, i.e., an array of meandering streamwise vortices, the Weissenberg number was increased up to 2000 (where the relaxation time was normalized by the wall speed and the kinematic viscosity).
We observed that, in the viscoelastic flows, the secondary flow (wall-normal and spanwise velocity fluctuations) are suppressed while the streamwise component is maintained, and the wavy roll cells are modulated into streamwise-independent straight roll cells. 
The kinetic energy transports among the mean shear flow, Reynolds stresses due to the roll cell, and viscoelastic stress are investigated in the non-turbulent background. We further discussed the modulation mechanism and its relevance to the drag-reduction phenomenon in the viscoelastic wall turbulence. 
\end{abstract}

\vspace{2pc}
\noindent{\it Keywords}: DNS, Drag reduction, Rotating plane Couette flow, Viscoelastic fluid

\maketitle

%%%%%%% New Section %%%%%%%%%%%%%%%%%%%%%%%%%%%%%%%%%%%%%%%%%%%%%%%%%%%%%%%%%%%%
\section{Introduction}

An improved understanding of viscoelastic, wall-bounded shear flow is essential for both the purposes for flow control as well as developing enhanced numerical and physical modeling with respect to the polymer-induced drag reduction phenomenon.
It is common knowledge, and may be known as the Toms effect, that polymer (or surfactant) additives in water significantly reduce the turbulent frictional drag in wall-bounded shear flows \cite{Lumley69,Zakin98,White08}. 
Such an additive solution exhibits viscoelasticity giving rise to profound modulations of the flow dynamics in wall turbulence as well as in laminar flows. 
Even in the laminar case, the modulation is generally difficult to be predicted without simulations or experiments and its relevance to the drag-reducing effect has been obscure for many years.
Although the simple linear properties of the viscoelastic fluid may be determined in simple shear or extensional laminar flows (that can be found in a rheometer), a non-linear complicated modulation induced by viscoelasticity needs to be determined while in elasto-inertial turbulence \cite{Groisman00,Samanta13,Dubief13}.
Understanding the mechanism of the drag reduction is still an important issue, even from the practical viewpoint, since a prediction of the drag-reducing effect would assist in improving the performance of fluid controls in relevant engineering situations. 
For instance, a drag reducer of surfactant has been applied as an energy-saving transport medium in oil pipelines and residential heating devices \cite{Takeuchi12}.
As a promising way to control turbulence, the polymer/surfactant additive can provide significant energy savings of up to about 80\%, which indicates a very high efficiency compared to other methods such as riblets, wall oscillations, and opposition control \cite{Gyr95,Frohnapfel12}.
Therefore, the mechanisms of drag reduction and turbulence modulation in the viscoelastic fluid flow are of great scientific and practical interest, attracting many researchers. 

Although the first theory of drag reduction goes back at least as far as \cite{Lumley69} or \cite{deGennes90}, this field has progressed with the aid of direct numerical simulation (DNS) after the first DNS using a FENE-P viscoelastic fluid model by \cite{Sureshkumar97}.
As in this pioneering work, other previous DNS studies have focused on canonical systems; for instance, the fully-developed turbulent channel flows \cite{dubief04,Yu04a,Yu04b,Housiadas03,Housiadas05,Li07,Thais13}, the minimal channel \cite{Xi10}, and the spatially-developing boundary layer \cite{Tamano07,Tamano09}. 
More recent works have also investigated the instabilities in viscoelastic fluid \cite{Kim08,Zhang13,Agarwal14,page14,Biancofiore17}. 
\cite{Kim08} tracked the evolution of an initially-isolated vortex, and found that the elasticity acting on the streamwise structures also suppressed the autogeneration of new vortices.
\cite{Biancofiore17} studied the secondary instability of nonlinear streaks, and reported that the polymer torque would change its role on the stabilization or destabilization of the streaks depending on their evolutional stages and amplitudes.
For the primarily weak streaks, the resistive polymer torque opposes the streamwise vorticity and hinders breakdown to turbulence.
They pointed out its relevance to the hibernating state that is prolonged at higher Weissenberg numbers.
During the hibernating state, the streaks remain stable for a relatively long time, which was observed through the minimal-channel DNS by \cite{Xi10}.
We also note a low-dimensional model study by \cite{Roy06}: their model does not consider the wall and revealed the effects of elasticity on a self-sustaining process of the coherent wavy streamwise vortical structures underlying wall-bounded turbulence.

For understanding the drag-reducing mechanism, one of the difficulties in directly studying the viscoelastic wall turbulence as done in many earlier studies is that, as the dynamics of the Newtonian wall turbulence itself is highly complex, it is difficult to interpret the effect by viscoelasticity from a change in flow structures. 
Generally, in the Newtonian wall turbulence, longitudinal structures such as quasi-streamwise rolls (or vortices) and low-speed streaks interact with each other giving rise to a self-sustaining process of the turbulent motions~\cite{waleffe97}. 
This process would be disturbed by addition of viscoelasticity, as examined by \cite{dubief04}, but the wide range of energy spectra of vortices including other-oriented ones renders the analysis complicated. 
As the elasticity itself may induce other kinds of instabilities, it is difficult to discriminate the influence of viscoelasticity from the self-sustaining process. 

Here, we propose to investigate viscoelastic plane Couette flow with spanwise system rotation (viscoelastic RPCF) as an alternative way to systematically reveal the viscoelastic effects on coherent streamwise-elongated vortical structures. 
In the Newtonian RPCF, 17 kinds of flow states appear depending on the Reynolds number and the system rotation rates due to linear instabilities by the Coriolis force, as experimentally observed by \cite{tsuka10}. 
As the laminar RPCF with anticyclonic system rotation exhibits well-organized roll cells with a variety of forms, such as two-dimensional straight roll cells and three-dimensional wavy roll cells, the laminar RPCF can be a good test case for studying how viscoelasticity modulates the roll cells. 
In particular, the wavy roll cell is characterized by its steady streamwise-\emph{dependency} and this fact allows us to systematically determine the viscoelastic effect on the instability of longitudinal vortex.
In present study, we investigate the laminar RPCF at $(\Rew, \Omega)=(100,10)$ for different Weissenberg numbers (the definitions of these parameters are given in the later section), where steady and three-dimensional wavy roll cells appear in the Newtonian fluid case, and thereby examine how viscoelasticity modulates the streamwise-dependent coherent vortical structures. 

This paper is structured as follows: Section 2 describes the flow configuration and governing equations. In Section 3, the numerical procedures are provided. 
Section 4 shows results that the streamwise-dependent wavy roll cells are modulated to streamwise-independent two-dimensional roll cells at moderate Weissenberg numbers. 
We discuss the onset of the steady streamwise-\emph{independency} in the viscoelastic flow and the relevance of the present results to the drag-reducing turbulent flow.

%%%%%%% New Section %%%%%%%%%%%%%%%%%%%%%%%%%%%%%%%%%%%%%%%%%%%%%%%%%%%%%%%%%%%%
\section{Flow configuration and governing equations}

%%%%%%% Figure 1 %%%%%%%%%
\begin{figure}
\begin{center}
	\includegraphics[scale=0.4]{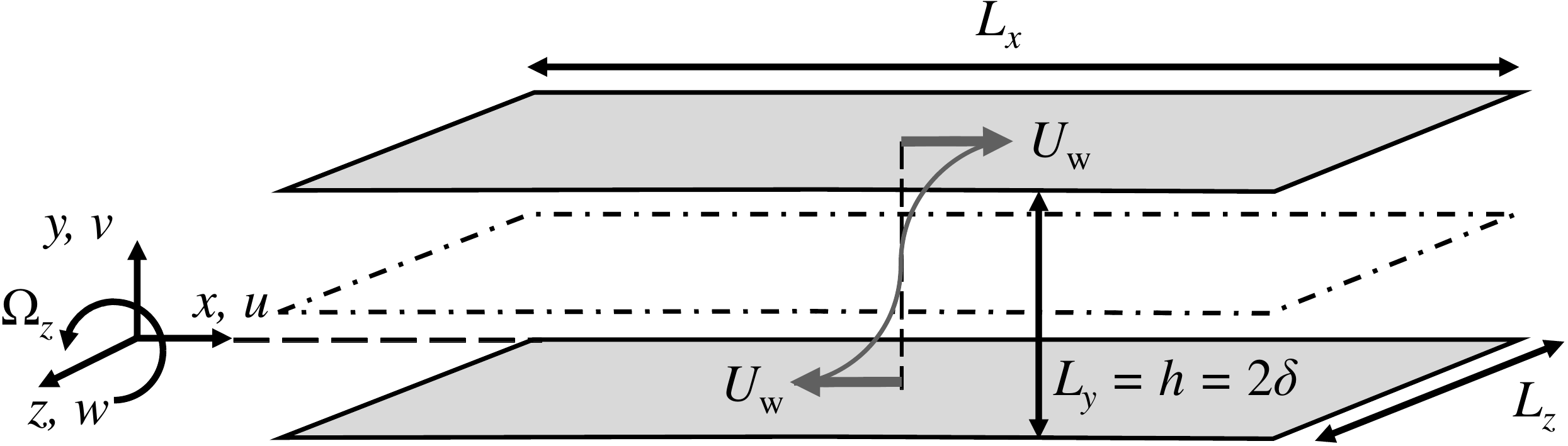}
	\caption{Configuration of rotating plane Couette flow.}
	\label{fig:RPCF}
\end{center}
\end{figure}

The rotating plane Couette flow (RPCF) we considered in the present DNS is defined in \Fref{fig:RPCF}. 
The flow is driven by two walls that are spatially separated by a distance $h$ and translating in opposite directions with the same speed $U_\mathrm{w}$, and such a plane Couette flow is subject to a spanwise system rotation at an angular velocity $\Omega_z$. 
The Cartesian coordinates are defined at an arbitrary streamwise and spanwise position on the bottom wall, and the $x$-, $y$-, and $z$-axes are taken in the streamwise, wall-normal, and spanwise directions, respectively. 

The governing equations are the incompressible continuity and non-Newtonian momentum equations in non-dimensional form:
\begin{equation}
   \pda{u_i}{x_i} = 0, 
   \label{eq:cont}
\end{equation}
\begin{equation}
   \pda{u_i}{t}+u_j^\ast \pda{u_i}{x_j} = -\pda{p}{x_i} + \frac{\beta}{\Rew} \pd{^2 u_i^\ast}{{x_j^\ast}^2}  - \frac{\Omega}{\Rew} \epsilon_{i3k} u_k^\ast + \frac{1- \beta}{\Wew} \pd{c_{ij}}{x_j^\ast},
   \label{eq:NS}
\end{equation}
where $u_i$, $p$, and $t$ are the velocity, pressure, and time. 
The superscript $^\ast$ stands for the quantities normalized by the wall speed $U_\mathrm{w}$ and/or the half channel height $\delta$. 
Here it should be noted that the pressure $p^\ast$ is defined subtracting the hydrostatic pressure $P$, which includes the centrifugal acceleration, as $p^\ast = (p - P)/\rho U_\mathrm{w}^2$. 
The Reynolds number and the rotation number are defined as ${\rm Re}_{\rm w}=U_{\rm w}\delta/\nu$ and $\Omega=2\Omega_z\delta^2/\nu$, respectively, where $\nu$ is the kinematic viscosity of the solution at zero shear rate. 
The third term in the right-hand side of \Eref{eq:NS} is the Coriolis force term with the Levi-Civita symbol $\epsilon_{ijk}$.

The last term in \Eref{eq:NS} is the viscoelastic force term, where $\beta$ is the ratio of the Newtonian solvent viscosity $\mu_\mathrm{s}$ to the total viscosity of the solution $\mu_\mathrm{s}+\mu_\mathrm{a}$ ($\mu_\mathrm{a}$ is the additive viscosity at zero shear rate), and $\Wew$ is the Weissenberg number $\Wew=U_\mathrm{w}^2 \lambda / \nu$, which represents the ratio of the relaxation time of the additive to the viscous time scale of the flow. 
The value of $1-\beta$ is a measure of the additive concentration, 
and we assume in this study that $\beta$ is constant in space and time. 
The non-dimensional viscoelastic stress tensor $c_{ij}$ is defined based on the conformation stress tensor $\tau_{ij}$ as $c_{ij}=\tau_{ij} \lambda / \mu_\mathrm{a} + \delta_{ij}$ ($\delta_{ij}$ is the Kronecker's delta). 
Equations~\eref{eq:cont} and \eref{eq:NS} are solved together with the constitutive equations of $c_{ij}$, 
for which we adopt the Giesekus model:
\begin{eqnarray}
 \pd{c_{ij}}{t^\ast}  &+ \pd{u_m^\ast  c_{ij}}{x_m^\ast} -\pd{u_i^\ast}{x_m^\ast} c_{mj} 
          -\pd{u_j^\ast}{x_m^\ast} c_{mi} \nonumber \\
          &+\frac{\Rew}{\Wew} \left[ c_{ij} -\delta_{ij} + \alpha (c_{im} - \delta_{im})(c_{mj} - \delta_{mj}) \right]= 0,
          \label{eq:Giesekus}
\end{eqnarray}
where $\alpha$ is the mobility factor and fixed at $\alpha=0.001$, which specific value was employed in previous studies \cite{Yu04b,tsuka11}. 
The Giesekus viscoelastic model considers the non-linearity by the last term with $\alpha$, which is absent in the other types of FENE-P and Oldroyd-B model. 
\cite{Yu04b} confirmed some reasonability between the Giesekus-model prediction and measured rheological properties of drag-reducer solution of low concentration surfactant. 
As \cite{Tamano07,Tamano09} demonstrated that three the Oldroyd-B, FENE-P, and Giesekus models qualitatively agreed in simulating a drag-reducing turbulent flow, we may consider that the use of different models would not qualitatively change the conclusions of the present study.

%%%%%%% New Section %%%%%%%%%%%%%%%%%%%%%%%%%%%%%%%%%%%%%%%%%%%%%%%%%%%%%%%%%%%%
\section{Numerical procedures}
We adopted the finite difference method for the spatial discretization. 
The forth-order central difference scheme was used for the $x$- and $z$-directions, while the one with second-order accuracy was adopted in the wall-normal ($y$-) direction. 
For the time integration, the second-order Crank-Nicolson and the second-order Adams-Bashforth schemes were used for the wall-normal viscous term and the other terms, respectively. 
As for the boundary condition, the periodic boundary conditions were imposed in the $x$ and $z$ directions and the non-slip condition was applied on the walls.

In the present numerical code, the minmod flux-limiter scheme was employed for the convective term in \Eref{eq:Giesekus} to stabilize the calculation that would often encounter a high-Weissenberg-number problem.
The performance of this scheme on numerical stability and accuracy with respect to drag-reducing turbulent channel flow was confirmed to be better than a local artificial-diffusion scheme \cite{Yu04b}. 

In this study we focus on the RPCF at $\Rew=100$ and $\Omega=10$, where in the Newtonian case three-dimensional wavy roll cells are observed~\cite{tsuka10}, and investigate four different Weissenberg number cases; Newtonian fluid case, $\Wew = 1000$, 1500, and 2000. The viscosity ratio $\beta$ is fixed at 0.8 for all cases. 
The computational domain size ($L_x \times L_y \times L_z$) is $7.5h \times h \times 3h$, so that one streamwise and spanwise wavelength of the 3D roll cells experimentally observed by \cite{tsuka10} can be captured in the domain.  
The grid number was $128 \times 128 \times 128$ with non-uniform grids in the $y$-direction; the grid resolutions were confirmed to be fine enough.
The simulations started with the linear velocity profile of the `non-rotating' laminar plane Couette flow, and the data were collected after the roll-cell structure fully developed. 

\begin{figure}[t]
\begin{center}
	\begin{minipage}{0.495\hsize}
		\includegraphics[viewport=0 0 590 226, clip, width=1\hsize]{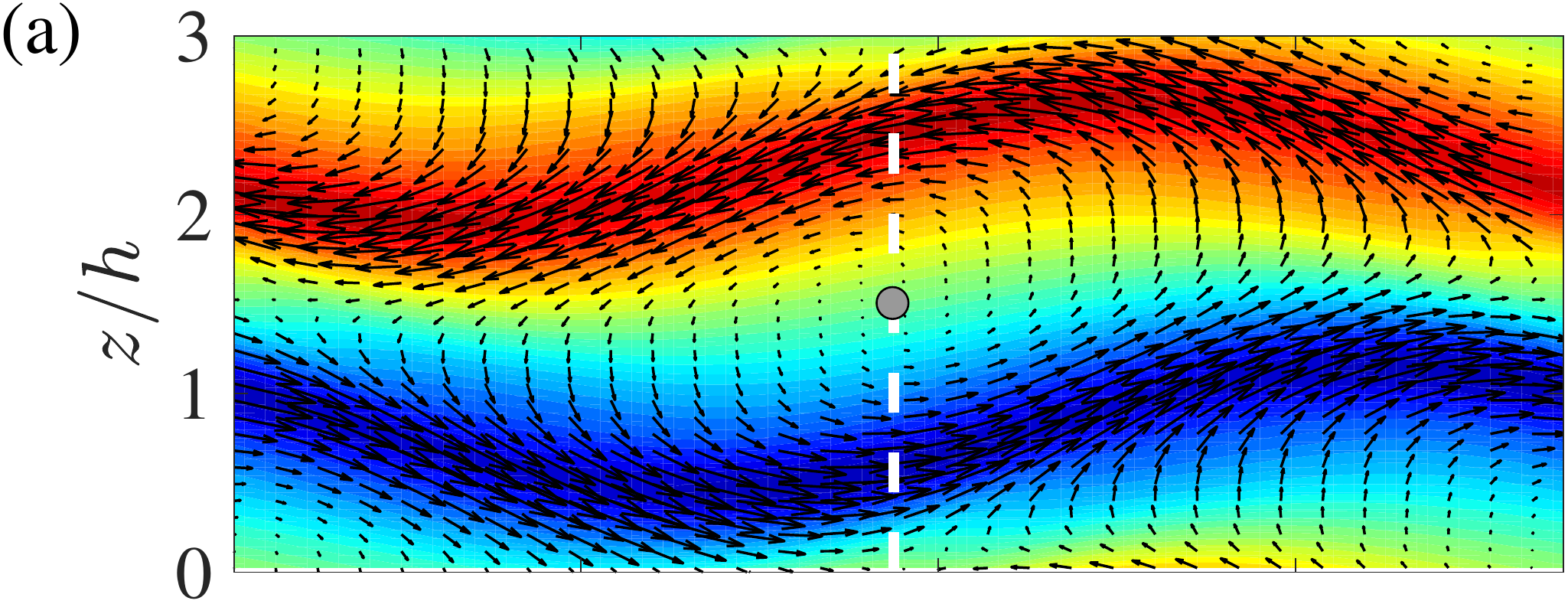}
	\end{minipage}
%	\vspace{-1.45\baselineskip}
	\begin{minipage}{0.495\hsize}
		\includegraphics[viewport=0 0 590 226, clip, width=1\hsize]{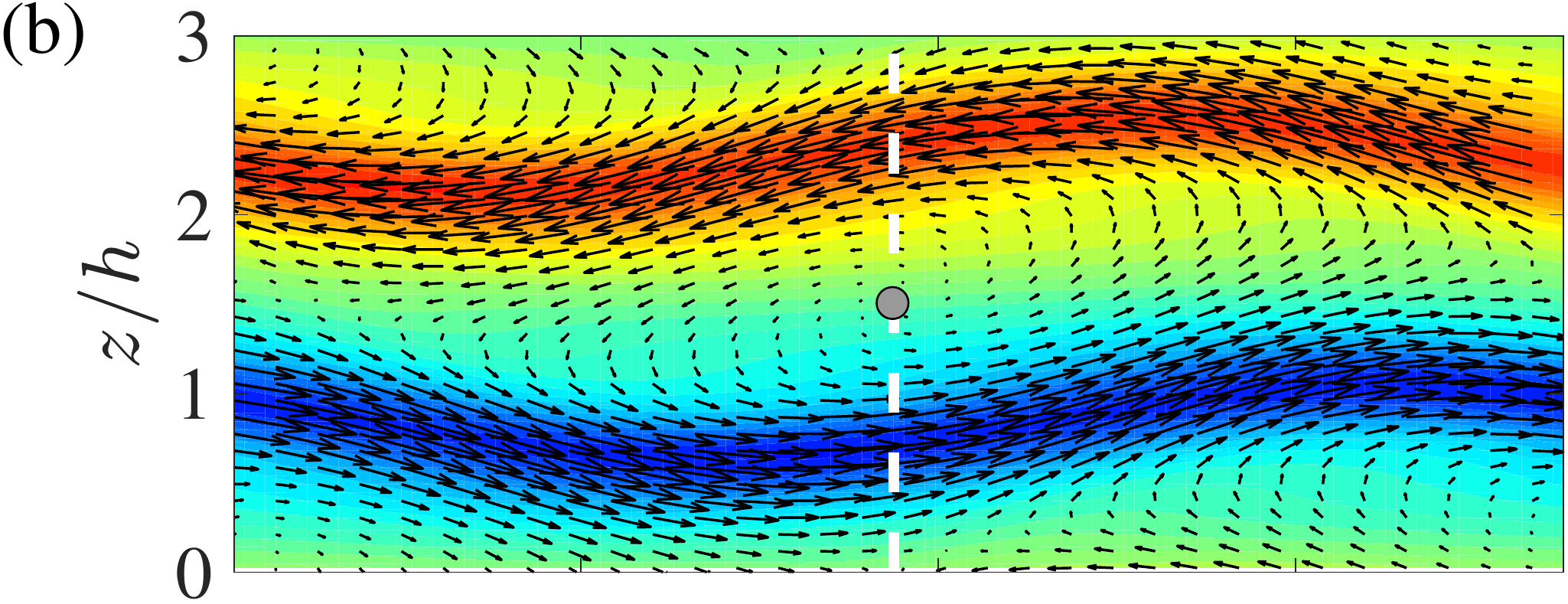}
	\end{minipage}
	\begin{minipage}{0.495\hsize}
		\includegraphics[width=1\hsize]{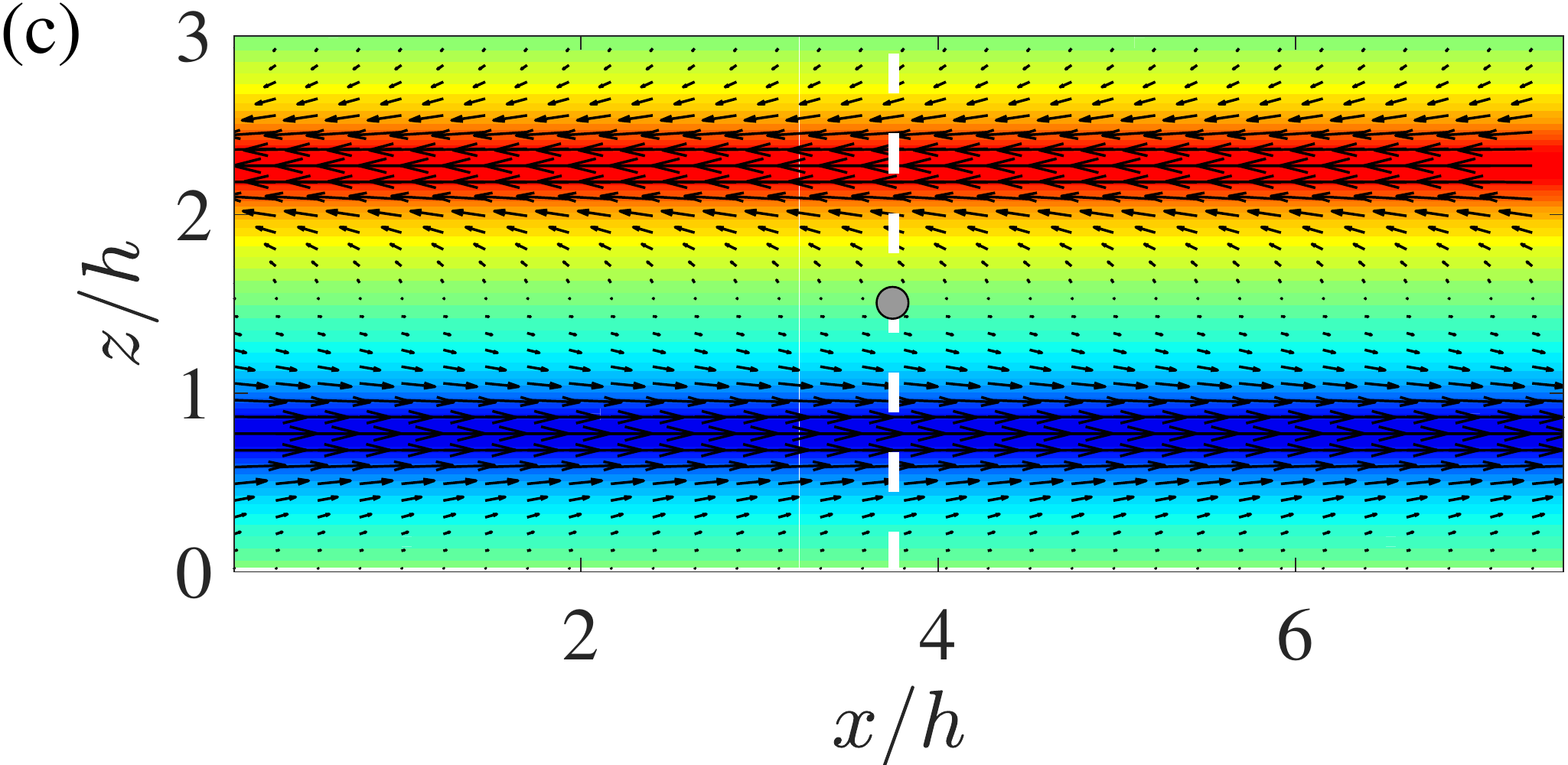}
	\end{minipage}
	\begin{minipage}{0.495\hsize}
		\includegraphics[width=1\hsize]{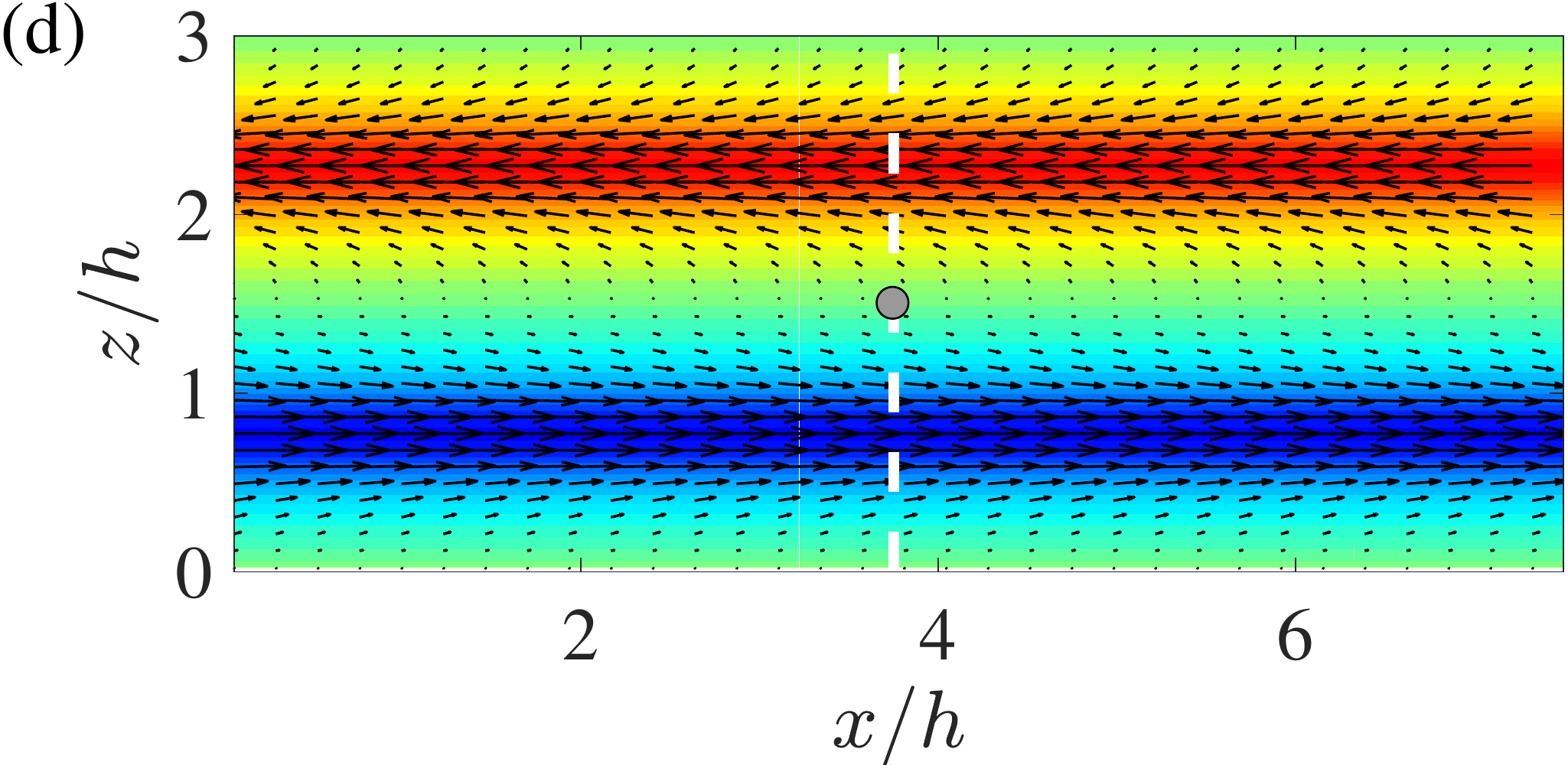}
	\end{minipage}
	\includegraphics[scale=0.4, clip, viewport=-60 -10 530 90]{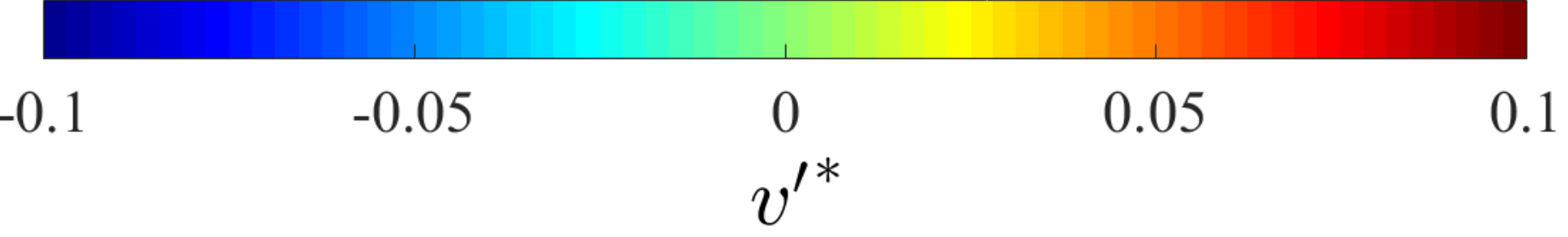}
	\vspace{-1\baselineskip}
	\caption{
Instantaneous velocity fields on $x$-$z$ plane at the channel center for (a) Newtonian fluid, (b) viscoelastic fluid with $\Wew=1000$, (c) $\Wew=1500$, and (d) $\Wew=2000$ cases. The contour and black arrows represent the wall-normal velocity fluctuation $v^\prime/U_\mathrm{w}$ and the in-plane velocity vector pattern, respectively. The vertical white dashed line in each panel indicates the streamwise positions of the cross-sectional planes on which the instantaneous velocity fields are visualized in \Fref{fig:zy_u}. 
The grey dot in each panel indicates the position around which the wall-normal torque presented in \Fref{fig:y_torque} is evaluated.}
	\label{fig:xz_v}
\end{center}
\end{figure}

%%%%%%% New Section %%%%%%%%%%%%%%%%%%%%%%%%%%%%%%%%%%%%%%%%%%%%%%%%%%%%%%%%%%%%
\section{Results}

In the present study, we define the mean velocities $\overline{u_i}$ as the values averaged in $x$- and $z$-directions at each $y$-position, and the velocity fluctuations $u_i^\prime$ as the deviation of the instantaneous velocities from the mean values: $u_i^\prime = u_i - \overline{u_i}$. 
As most of the flows observed in the present study are steady, the velocity `fluctuation' in this study is constant in time and only represents the spatial variation of velocity due to the secondary flow of the roll cells. 
Although such decomposition is usually adopted for turbulent flows, we also apply it for the current laminar flow with steady roll cells, since based on such definition of the fluctuating velocities the Reynolds-stress transport equations are easily derived, which allow us to evaluate the energy exchange between the mean conformation tensor $\overline{c_{ij}}$ and the flow kinetic energy.

\subsection{Instantaneous flow fields}

\begin{figure}[t]
\begin{center}
	\begin{minipage}{0.495\hsize}
	\includegraphics[viewport=0 0 679 226, clip, width=1\hsize]{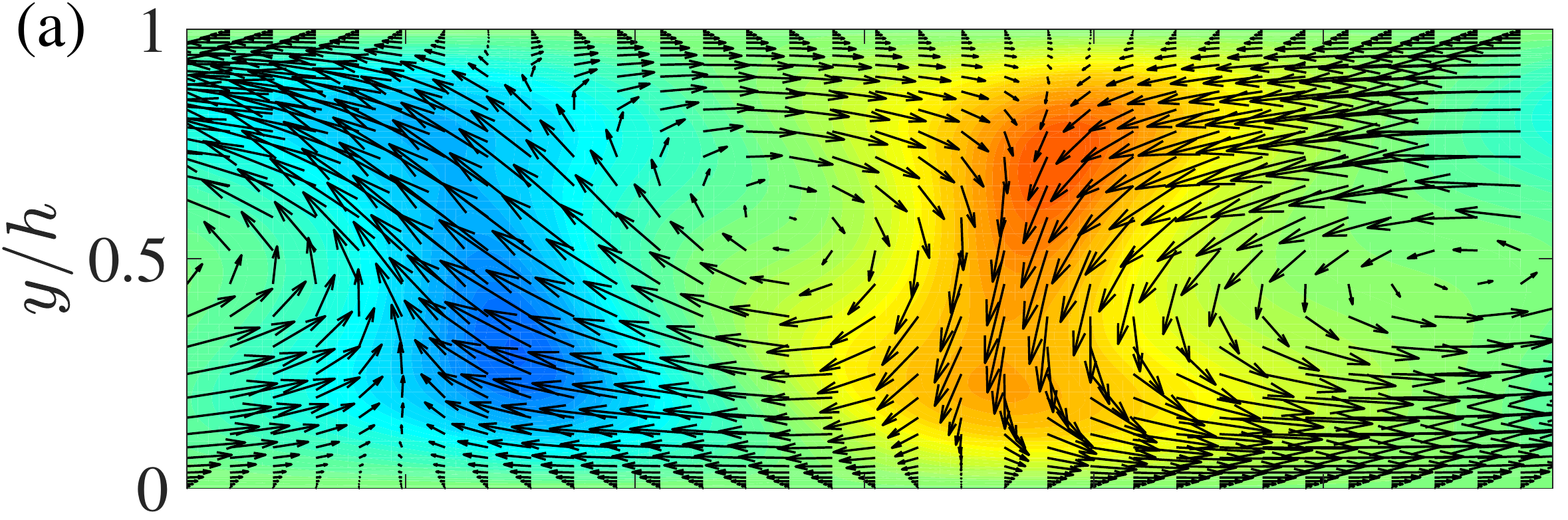}
	\end{minipage}
	\begin{minipage}{0.495\hsize}
	\includegraphics[viewport=0 0 679 226, clip, width=1\hsize]{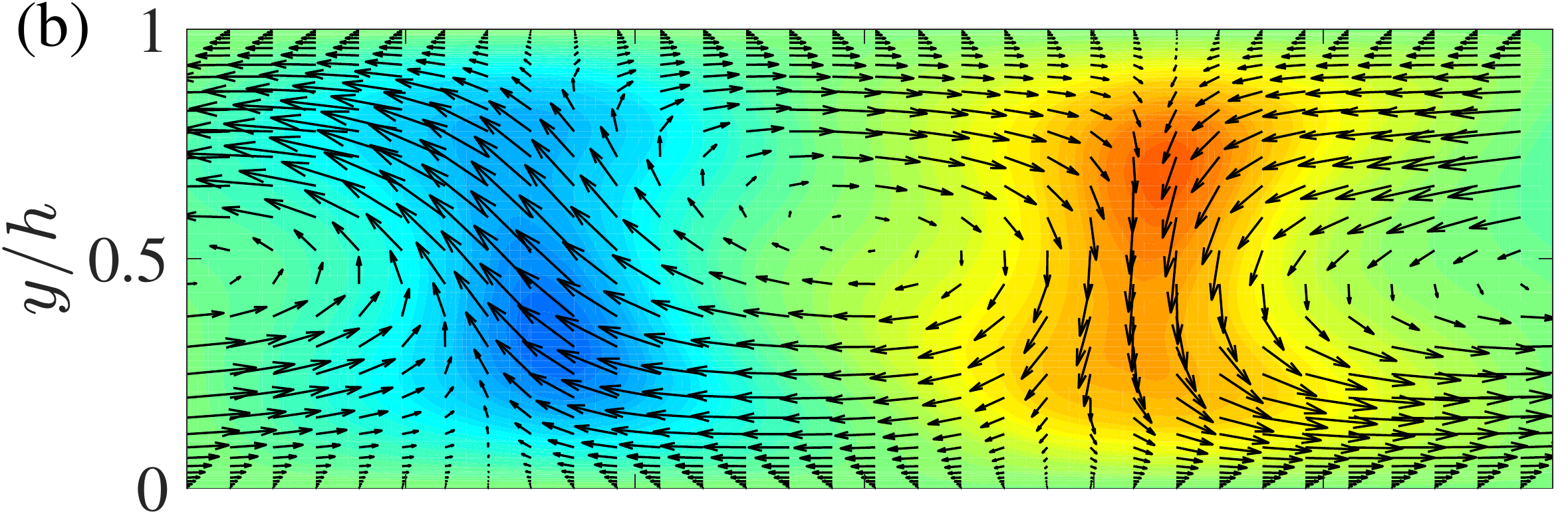}
	\end{minipage}
	\begin{minipage}{0.495\hsize}
	\includegraphics[width=1\hsize]{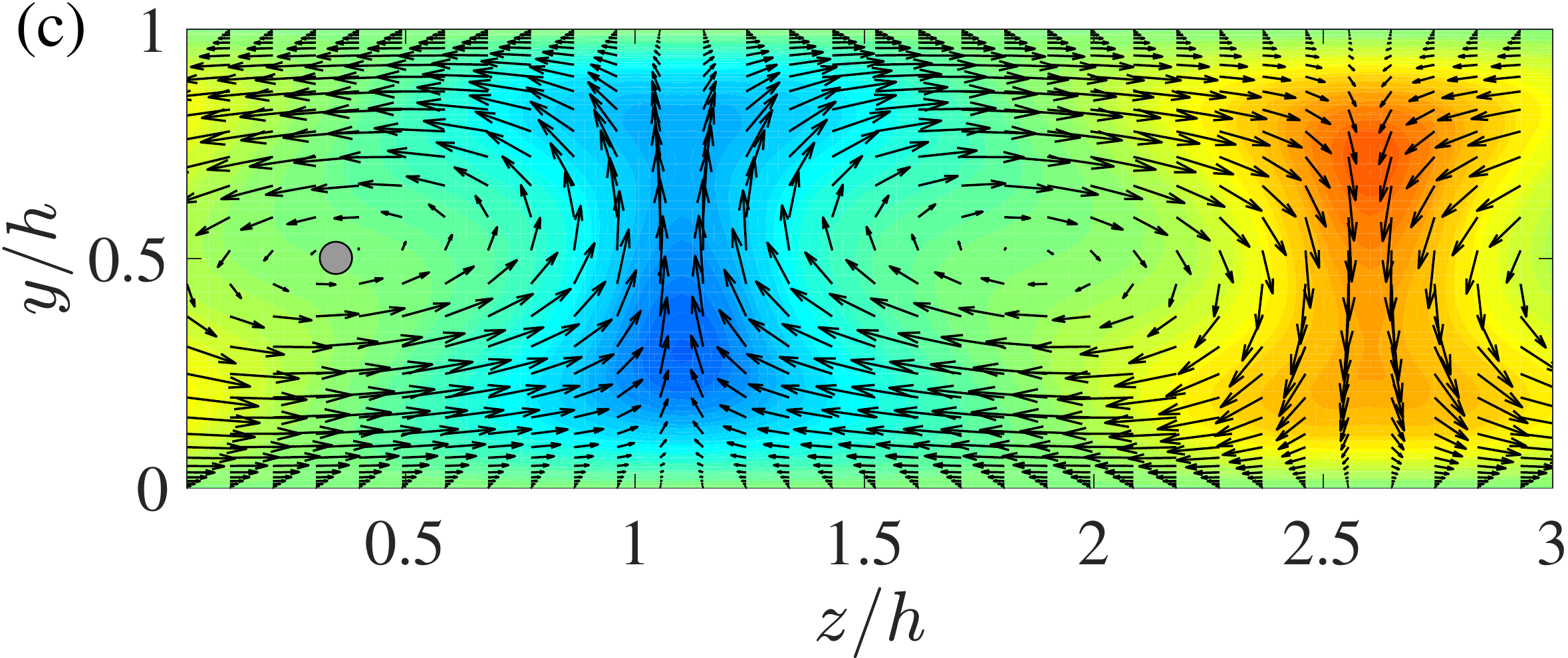}
	\end{minipage}
	\begin{minipage}{0.495\hsize}
	\includegraphics[width=1\hsize]{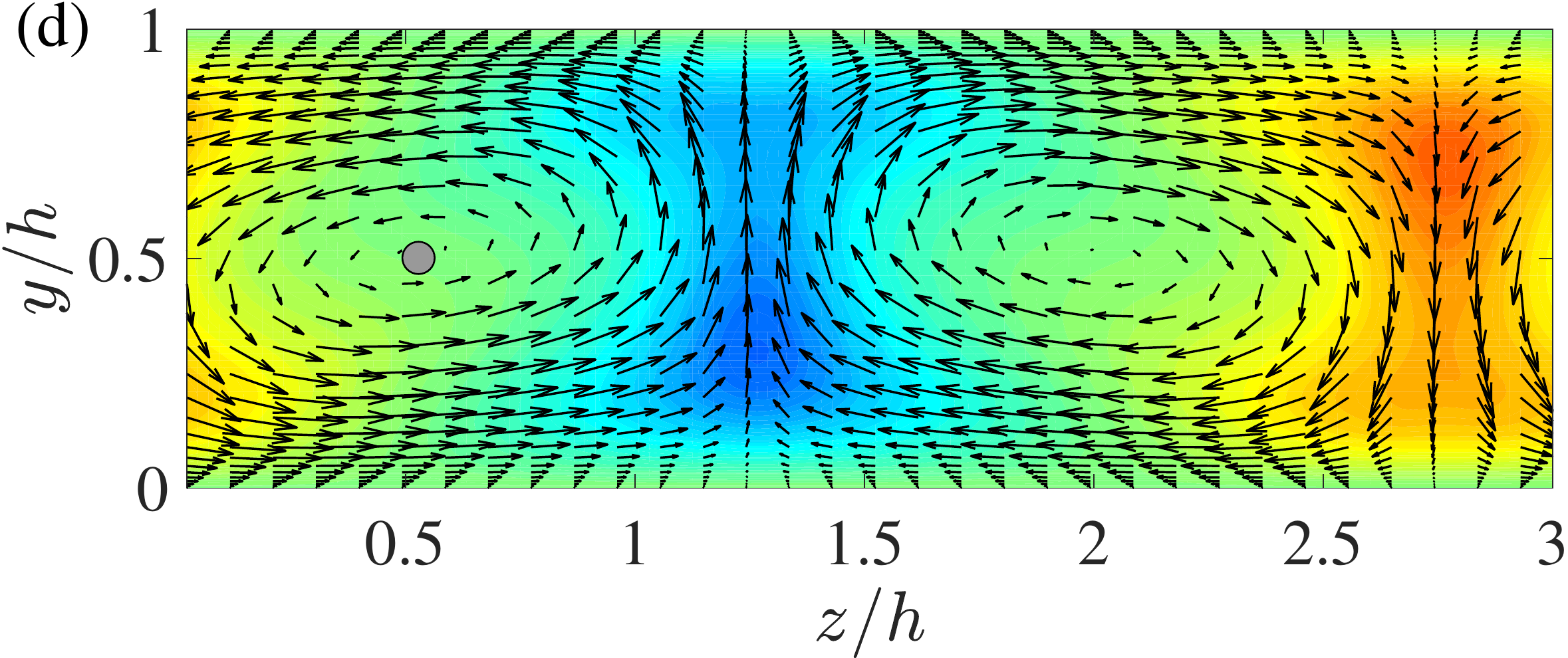}
	\end{minipage}
	\includegraphics[scale=0.4, clip, viewport=-30 0 623 80]{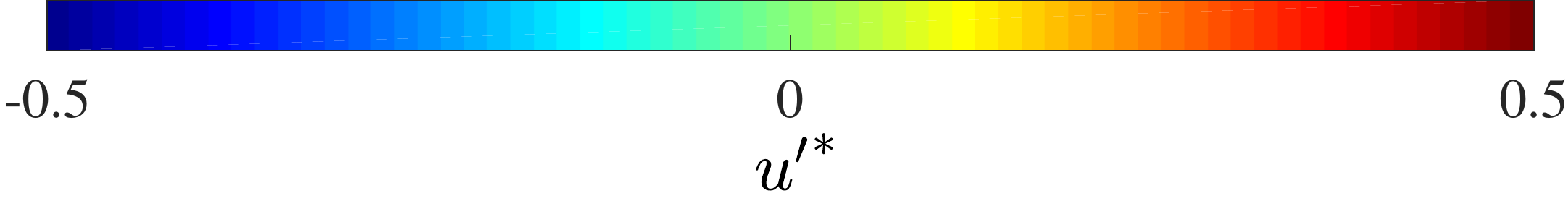}
	\vspace{-1\baselineskip}
	\caption{Instantaneous velocity fields on cross-sectional plane at the streamwise position indicated by the white dashed lines in \Fref{fig:xz_v} for (a) Newtonian, (b)$\Wew=1000$, (c) $\Wew=1500$, (d) $\Wew=2000$ cases. The contour and black arrows represent the streamwise velocity fluctuation $u^\prime/U_\mathrm{w}$ and the in-plane velocity vector pattern, respectively. 
The grey dots in (c) and (d) panel indicate the position around which the streamwise torque for the streamwise-independent roll cells presented in \Fref{fig:x_torque} is evaluated.}
\label{fig:zy_u}
\end{center}
\end{figure}

Figures~\ref{fig:xz_v} and \ref{fig:zy_u} present snapshots of the obtained instantaneous flow fields of the Newtonian/viscoelastic RPCF with different $\Wew$. 
\Fref{fig:xz_v} provides the velocity field on the channel central plane, while \Fref{fig:zy_u} displays the streamwise cross-sectional slice of the roll-cell structure at the streamwise positions indicated by the white dashed lines in \Fref{fig:xz_v}. 
%The velocity fluctuations shown in the figures are defined as the deviation from the mean values of each component: $u_i^\prime=u_i-\overline{u_i}$, where $\overline{u_i}$ is the mean velocities averaged in $x$- and $z$-directions and in time (but basically $\overline{v}=\overline{w}=0$ at any height). 
From the figures, one can observe that a pair of streamwise-elongated roll cells is captured in all cases: in \Fref{fig:xz_v}, the vectors exhibit positive and negative speed streaks in $x$, and they are associated with upward (red contour, $v^\prime > 0$) and downward (blue, $v^\prime < 0$) motions in $y$.
This roll cell and streak structure occurs spontaneously without any disturbance due to the linear instability of RPCF \cite{Lezius76}.
At the present Reynolds number, one may encounter a spanwise meandering of roll cells, as predicted by \cite{Nagata98}.
As shown in \Fref{fig:xz_v}(a), streamwise-dependent and wavy roll cells are actually confirmed in the Newtonian case, which is consistent with the experimental observation by \cite{Hiwatashi07,tsuka10}. 
As the Weissenberg number increases with fixed $\Rew$ and $\Omega$, such a wavy pattern of the roll cells is modulated, and at $\Wew=1500$ and 2000 the roll cells are two-dimensional and straight in the streamwise direction, as indicated in \Fref{fig:xz_v}(c) and (d). 
The flow at $\Wew=2000$ is not steady in time and slightly pulsates in terms of the magnitudes of the roll cell and positive/negative-speed streak.
However, we did not find any noticeable temporal change in the flow pattern, as discussed later.

Corresponding to the wavy pattern of the roll cells observed in the Newtonian case, the secondary flow given in \Fref{fig:zy_u}(a) is asymmetric with respect to the channel centerline, and a significant spanwise fluid motion that diagonally crosses between the neighboring vortices is observed. 
Comparing the viscoelastic cases with the Newtonian case, one can see that at $\Wew=1000$ the diagonal spanwise secondary flow is somewhat suppressed, and at the higher Weissenberg numbers where the roll-cell structure is modulated to the 2D and streamwise independent, the secondary flow pattern is symmetric in the wall-normal direction, as shown in \Fref{fig:zy_u}(c) and (d). 

%%%%%%% Figure 4 %%%%%%%%%
\begin{figure}[t]
\begin{center}
	\includegraphics[height=70mm]{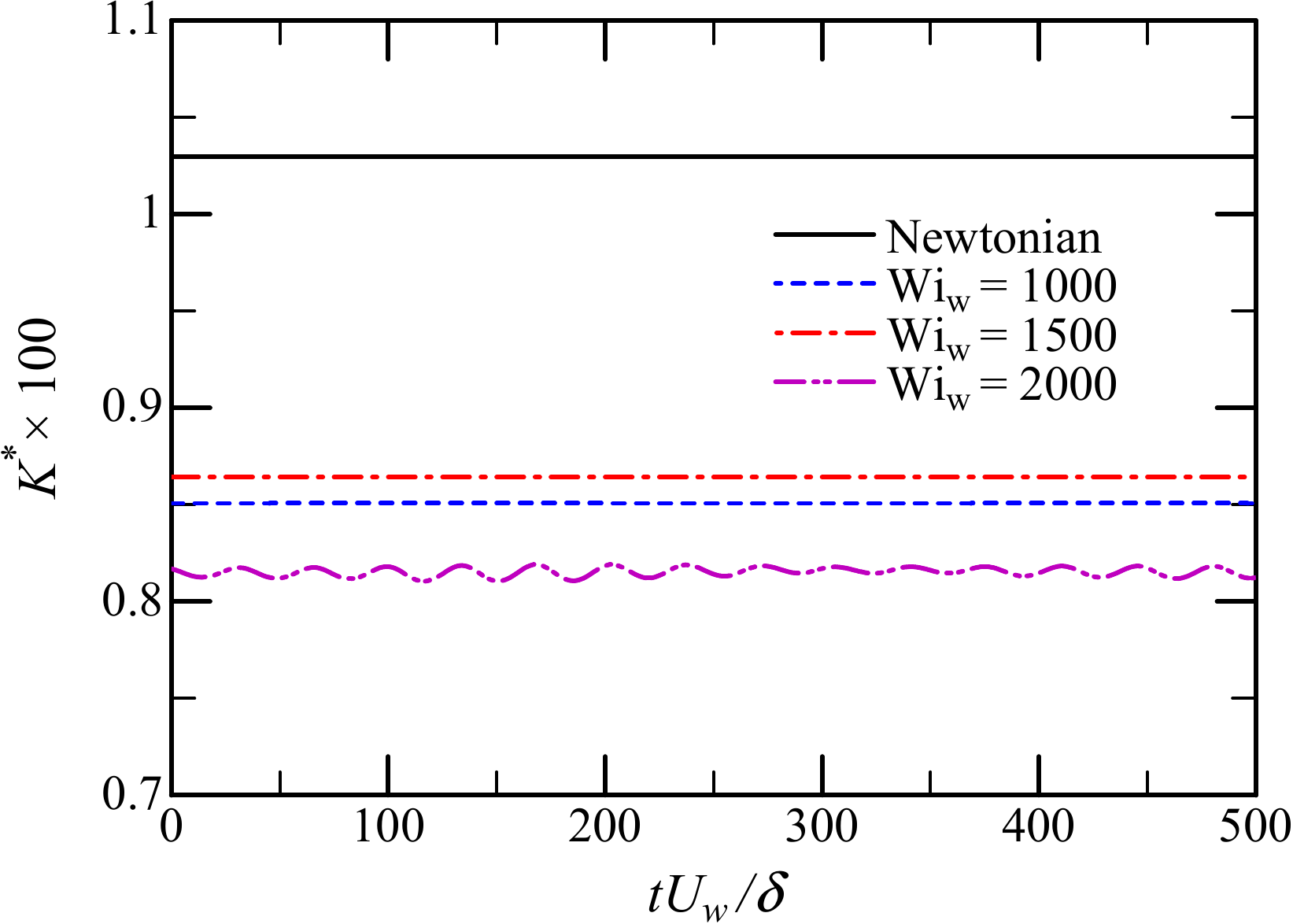}
	\caption{Temporal variation of volume-averaged kinetic energy $K$ scaled by $U_\mathrm{w}^2$. }
	\label{fig:k_t}
\end{center}
\end{figure}
%%%%%%% Figure4 %%%%%%%%%%%

To investigate the temporal behavior of these roll-cell structures, the volume-averaged kinetic energy is evaluated based on the fluctuating velocities as: 
\begin{equation}
      K=\frac{1}{L_x L_y L_z} \int^{L_x}_0 \int^{L_z}_0 \int^{L_y}_0 \frac{1}{2}({u^\prime}^2+{v^\prime}^2+{w^\prime}^2) {\rm d}x{\rm d}y{\rm d}z,
      \label{eq:K}
\end{equation}
and the time sequences are presented in \Fref{fig:k_t}, where $t=0$ corresponds to an arbitrary instance after the respective flow-pattern developments. 
The values of $K$ decrease noticeably in the viscoelastic fluids compared to the Newtonian case, as the diagonal spanwise secondary flow as well as the roll cell itself are suppressed in the higher $\Wew$ cases, as shown in \Fref{fig:zy_u}. 
The roll-cell structure is steady in the Newtonian and low-$\Wew$ viscoelastic cases, while the $\Wew=2000$ case indicates a certain periodic unsteadiness, indicating that the viscoelasticity may not only modulate the streamwise-dependency of the roll-cell structure, but also give rise to other kind instabilities. 
Such viscoelasticity-induced instabilities become more significant as the Weissenberg number increases, and may have some relevance to hibernating turbulence and the elasto-inertia turbulence \cite{nimura17,nimura18}. 
Further investigation on such viscoelastic-induced instabilities is, however, beyond the scope of the present study, and will be our future task.

\begin{figure}[t]
\begin{center}
	\includegraphics[height=55mm]{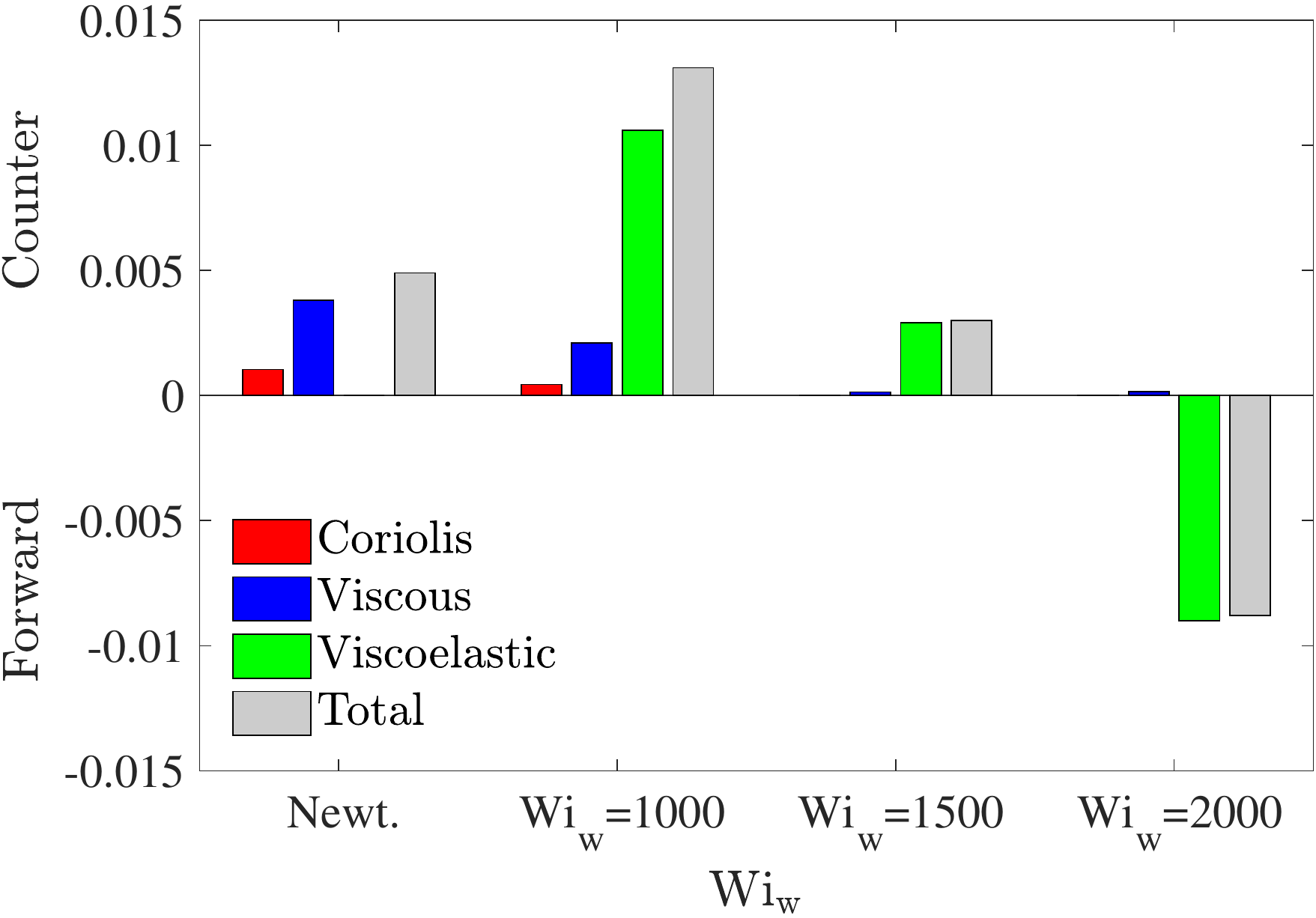}
	\caption{Torques acting on the wall-normal vorticity for the Newtonian case and the different Weissenberg number cases. The torque is determined with respect to the apparent center of wall-normal vortex motion in counter-clockwise direction, which is marked with a grey dot in each corresponding panel of \Fref{fig:xz_v}.}
	\label{fig:y_torque}
\end{center}
\end{figure}

In order to investigate how the viscoelastic force contributes to the modulation of the roll cells, the wall-normal torque acting on the roll cells at the channel center shown in \Fref{fig:xz_v} is evaluated as 
\begin{equation}
{\rm Torque}_y=\int\int  \left[ (z-z_c) F_x - (x-x_c)F_z \right] \mathrm{d}x \mathrm{d}z, 
\end{equation}
where the location of $(x_c,z_c)$ in each case is indicated by the grey circle in \Fref{fig:xz_v}. 
In \Fref{fig:y_torque}, the Coriolis, viscous, and viscoelastic contributions are compared, and the positive and negative torques indicate the favorable and against contribution to wall-normal vortex-like motion, respectively. 
As shown in the figure, the Coriolis and viscous torques counteract the local wall-normal vorticity in the Newtonian case. 
In the case of $\Wew=1000$ the viscoelastic torque also shows the contribution in the same way but the magnitude is significant compared to the Coriolis and viscous contributions. 
Such viscoelastic effect of stabilizing secondary instability of streaks is also observed in transient algebraic growth of optimal disturbance in the plane Couette flow (without system rotation) at a transitional Reynolds number \cite{Biancofiore17}. 
It is also shown in \Fref{fig:y_torque} that, as $\Wew$ increases, the positive viscoelastic torque decreases, and at $\Wew=2000$ the viscoelastic contribution turns to negative. 
It is interesting that between $\Wew=1500$ and $\Wew=2000$ the viscoelastic torque contributes in the opposite ways although the flow fields in the both $\Wew$ cases are quite similar, as shown in Figures~\ref{fig:xz_v} and \ref{fig:zy_u}.

\begin{figure}[t]
\begin{center}
	\includegraphics[height=55mm]{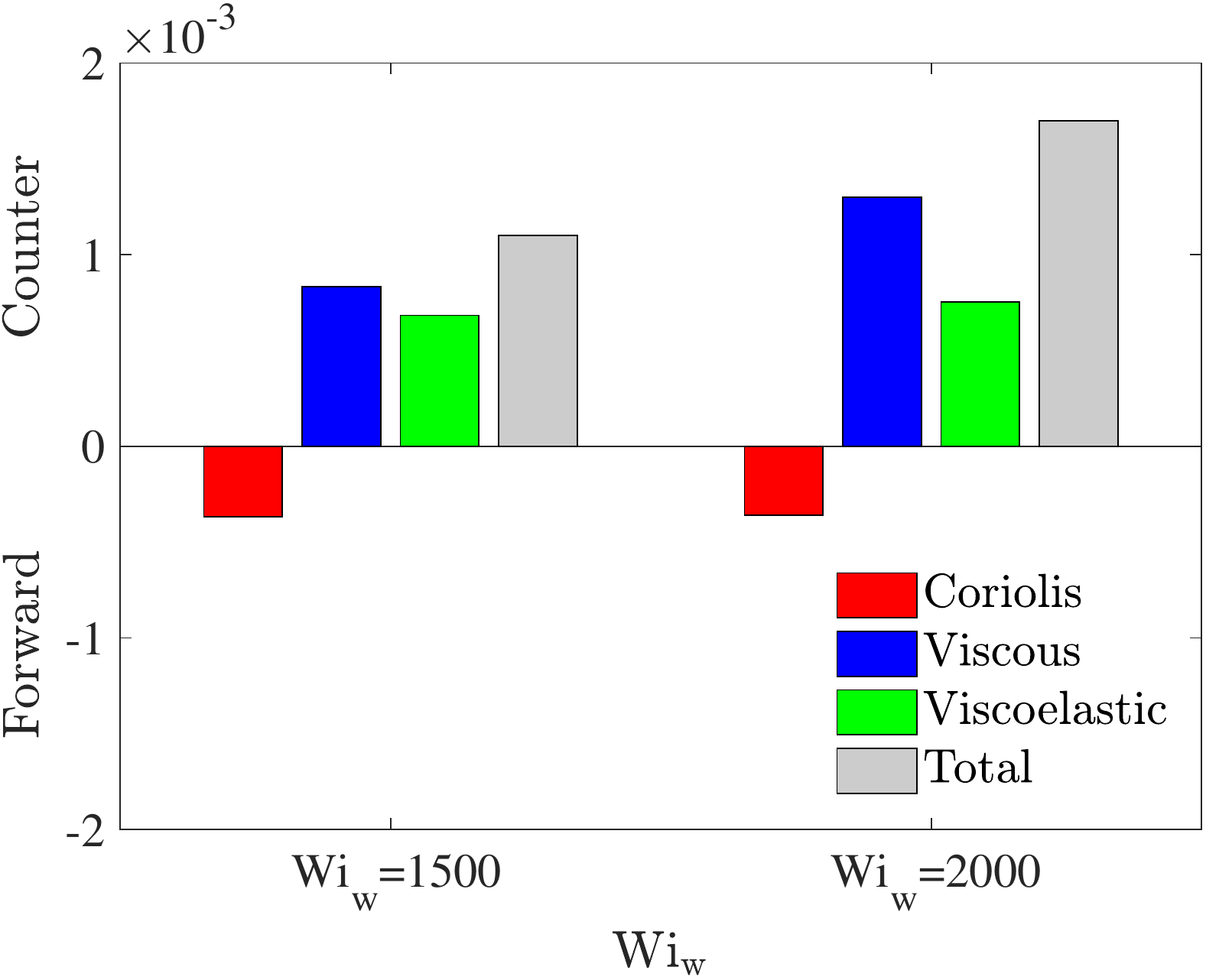}
	\caption{Torques acting on the two-dimensional roll cells at $\Wew=1500$ and $2000$. The torque is determined with respect to the center of streamwise-vortex rotating in counter-clockwise direction, which is marked with a grey dot in \Fref{fig:zy_u}(c) and (d).}
	\label{fig:x_torque}
\end{center}
\end{figure}

\Fref{fig:x_torque} compares the torque around the axis of the streamwise-independent roll cells observed at $\Wew=1500$ and $\Wew=2000$. 
In this figure, the torques around the point marked with a grey circle in Figures~\ref{fig:zy_u}(c) and (d) are compared. 
It is shown that in both cases the Coriolis force is acting favorably to the roll cell, while the viscous and viscoelastic forces exhibit counter torques and their magnitudes are almost unchanged. 
Such a tendency of viscoelasticity counteracting the secondary flow is also observed by \cite{page15,Biancofiore17}. 

\subsection{Mean flow profiles and velocity fluctuations}

%%%%%%% Figure 5  %%%%%%%%%
\begin{figure}[t]
\begin{center}
 \includegraphics[height=70mm]{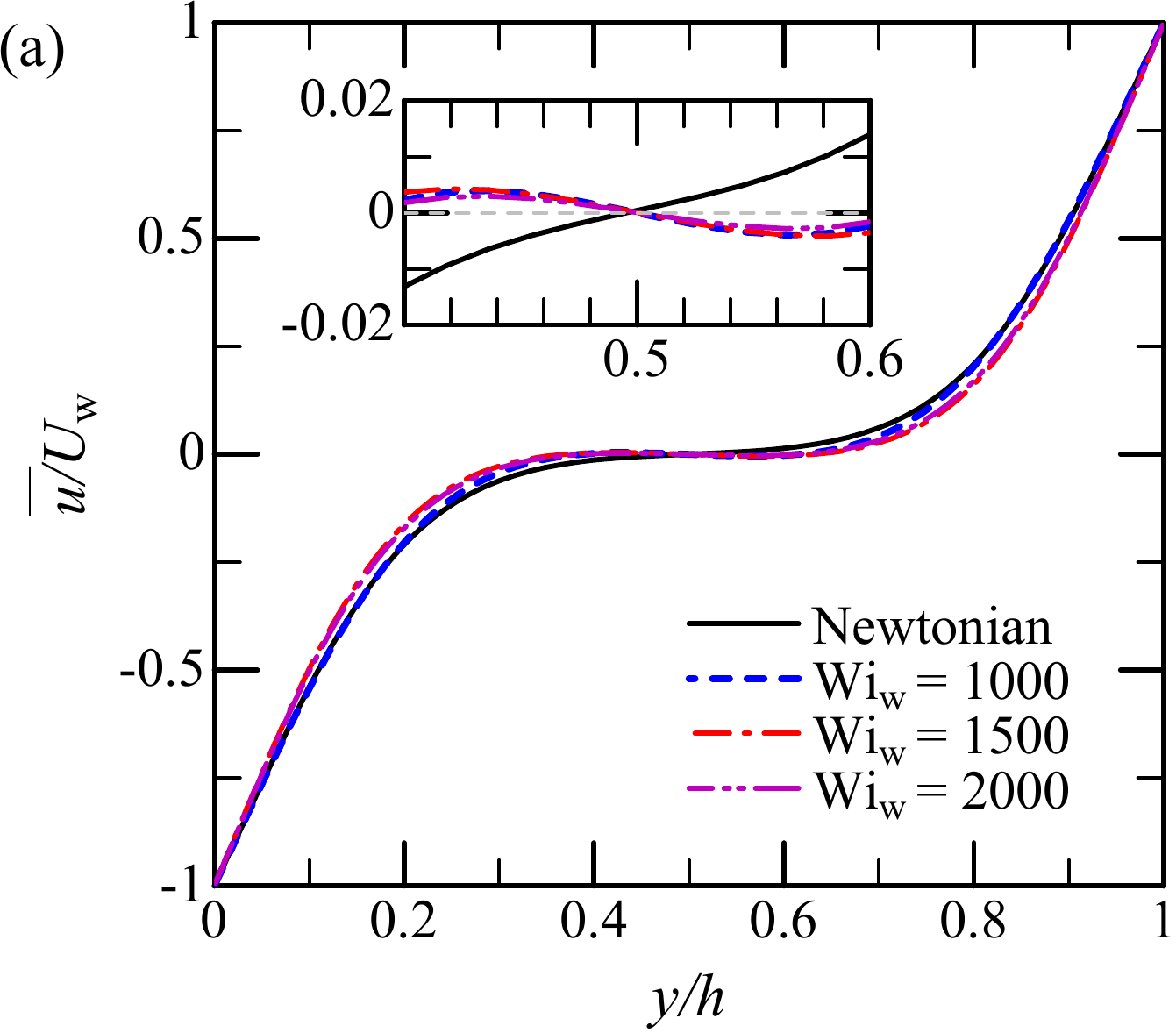} \hspace{1em}
 \includegraphics[height=70mm]{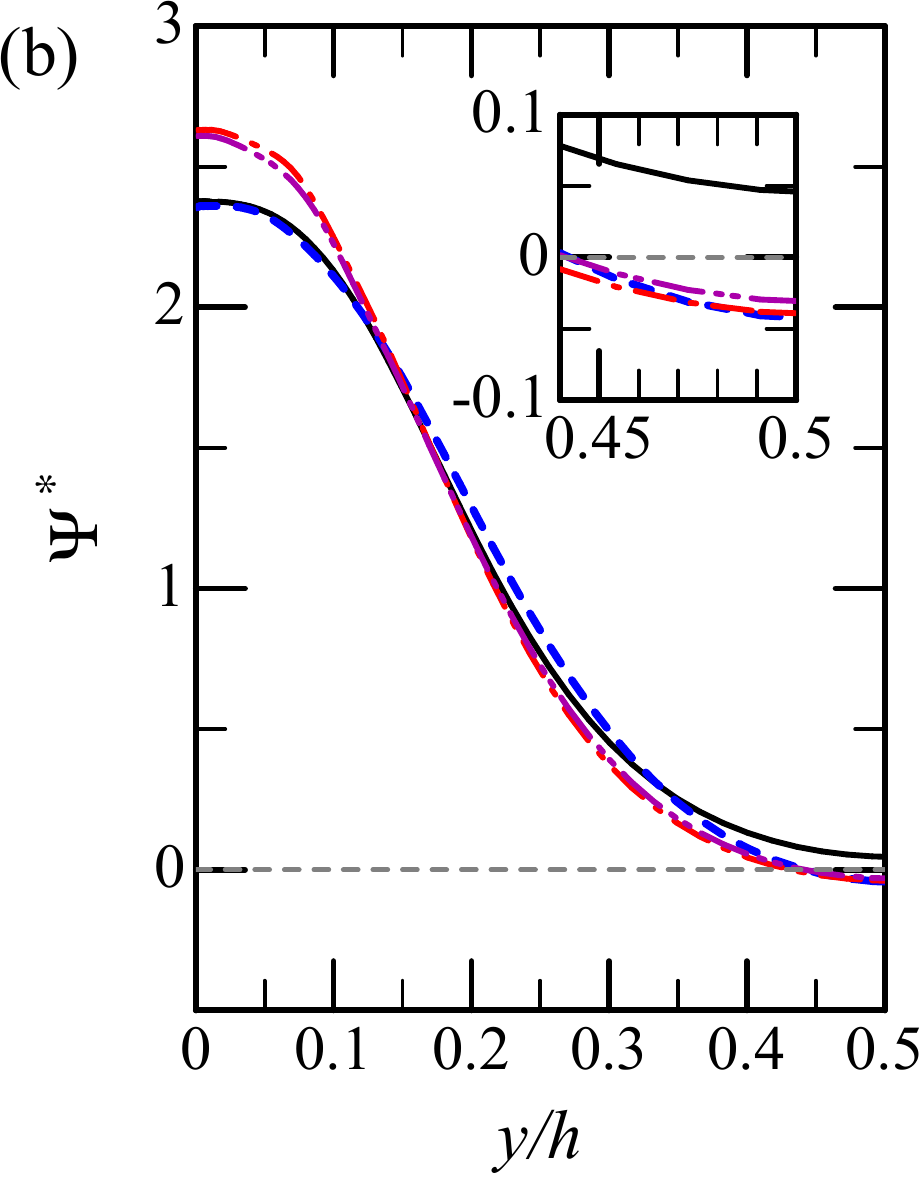}
 \caption{Profiles of (a) mean streamwise velocity $\overline{u}(y)$ and (b) mean velocity gradient $\Psi = {\rm d}\overline{u}/{\rm d}y$. They are scaled by $U_\mathrm{w}$ and $h$. Only half the channel is plotted in (b). Inserts are enlarged views with emphasis on the central region.}
 \label{fig:mean_u}
\end{center}
\end{figure}
%%%%%%% Figure 5 %%%%%%%%%%%

\Fref{fig:mean_u} presents the ensemble-averaged streamwise velocity and its wall-normal gradient. 
As shown in \Fref{fig:mean_u}(a), the mean velocity profiles are `turbulent-like' in all cases due to the additional momentum transport caused by the roll-cell structure. 

Comparing the profiles of the mean velocity gradient $\Psi$ in \Fref{fig:mean_u}(b), one can see that $\Psi$ values on the wall at $\Wew=1500$ and 2000 are about 15\% larger in those in the Newtonian and $\Wew=1000$ cases, whereas in the central region of the channel the $\Psi$ values at $\Wew=1500$ and 2000 are smaller. 
It is also noteworthy that in all the viscoelastic cases, the velocity gradient $\Psi$ is negative at the channel center. 
Such a negative mean velocity gradient at the channel center has also been observed in the Newtonian RPCF in turbulent flow regime~\cite{kawata16b,gai16}, but not at such low Reynolds numbers in the laminar flow regime. 
This tendency of the mean velocity gradient observed here indicates that the momentum transport across the channel is enhanced by the effect of the viscoelasticity in the higher $\Wew$ cases, where the non-wavy 2D roll cells forms, than in the lower $\Wew$ or Newtonian cases with 3D wavy roll cells. 

%%%%%%% Figure 6 %%%%%%%%%%%
\begin{figure}
\begin{center}
	\includegraphics[width=0.5\hsize]{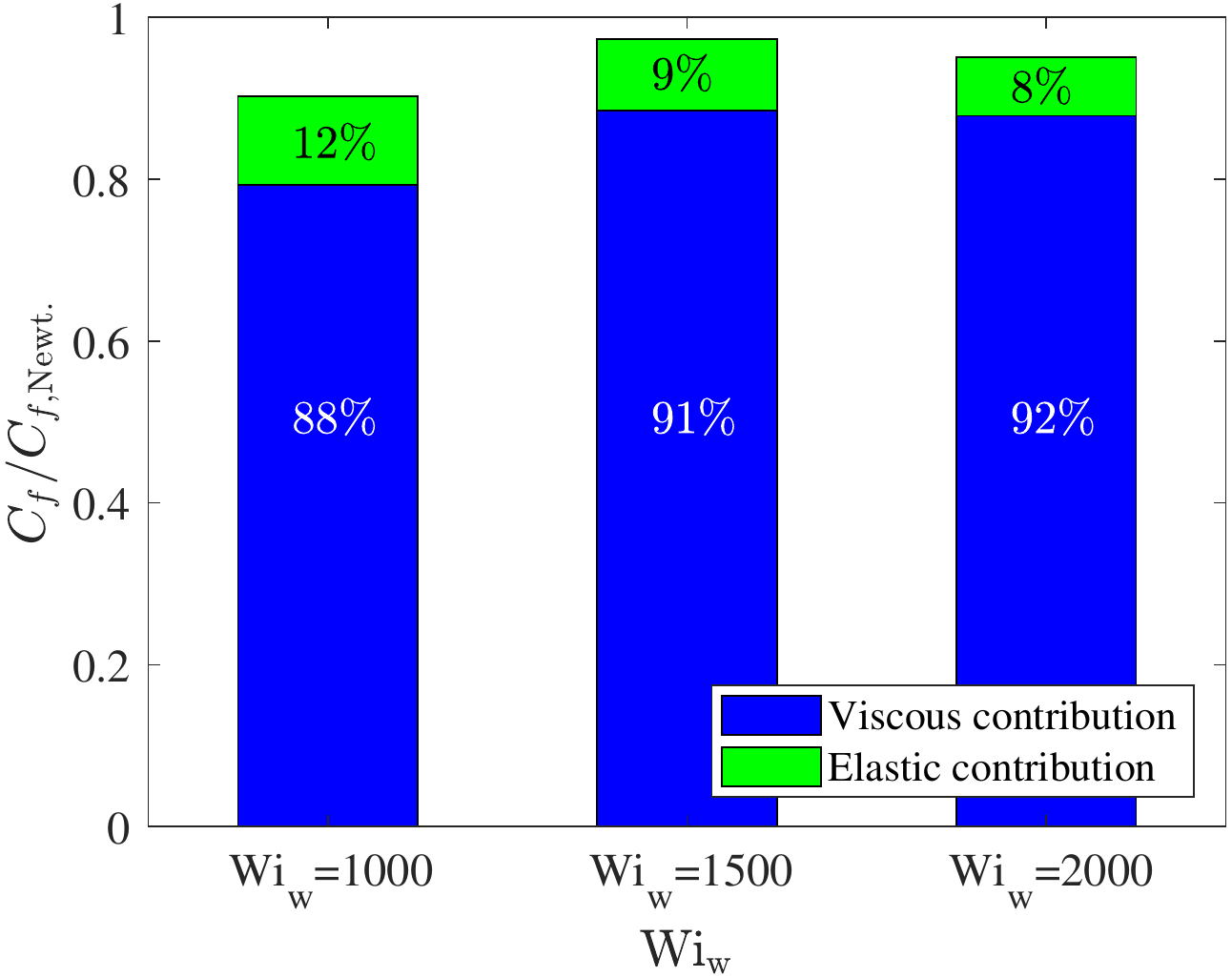}
	\caption{Variation in the skin friction coefficient and the contributions from the mean flow and the viscoelasticity at three different Weissenberg-number cases. The values of $C_f$ are normalized by the Newtonian value $C_{f, \mathrm{Newt}}$.}
	\label{fig:cf}
\end{center}
\end{figure}
%%%%%%% Figure 6 %%%%%%%%%%%

\Fref{fig:cf} shows the skin friction coefficient for different $\Wew$, defined as:
\begin{equation}
C_f \equiv \frac{2\tau_\mathrm{w}}{\rho U_\mathrm{w}^2}=2 \left( \left. \frac{\beta}{\Rew}  \frac{{\rm d} \overline{u}^\ast}{{\rm d}y^\ast}  \right|_{y=0} +  \frac{ 1-\beta}{\Wew} \left. \overline{c}_{12}  \right|_{y=0} \right), 
\label{eq:cf}
\end{equation}
where $\tau_\mathrm{w}$ is the wall shear stress and $\rho$ the density.
The first and second term in the right-hand side indicate the contribution from the viscous and the viscoelastic shear stress on the wall, respectively.  The values presented in \Fref{fig:cf} are normalized by the Newtonian case value $C_{f{\rm ,Newt.}}$, and the ratio of the viscous and viscoelastic contributions to the total $C_f$ are also presented in the figure. 
It is shown that the viscous contribution accounts for most of the total skin friction in all $\Wew$ cases, and the total skin friction decreases in all the viscoelastic cases compared to the Newtonian case. The $C_f$ decrease 10\%, 3\%, and 5\% in the $\Wew=$1000, 1500, and 2000 cases, respectively, and particularly in the $\Wew=1500$ and 2000 cases, the mean velocity gradient at the wall is significantly larger than the Newtonian case although the total $C_f$ is smaller. This is because $\beta=0.8$ is multiplied to the viscous contribution as shown by \Eref{eq:cf}, and even together with the elastic contribution, the total $C_f$ does not exceed the Newtonian value.   

%%%%%%% Figure 7 %%%%%%%%%
\begin{figure}
\begin{center}
 \includegraphics[height=70mm]{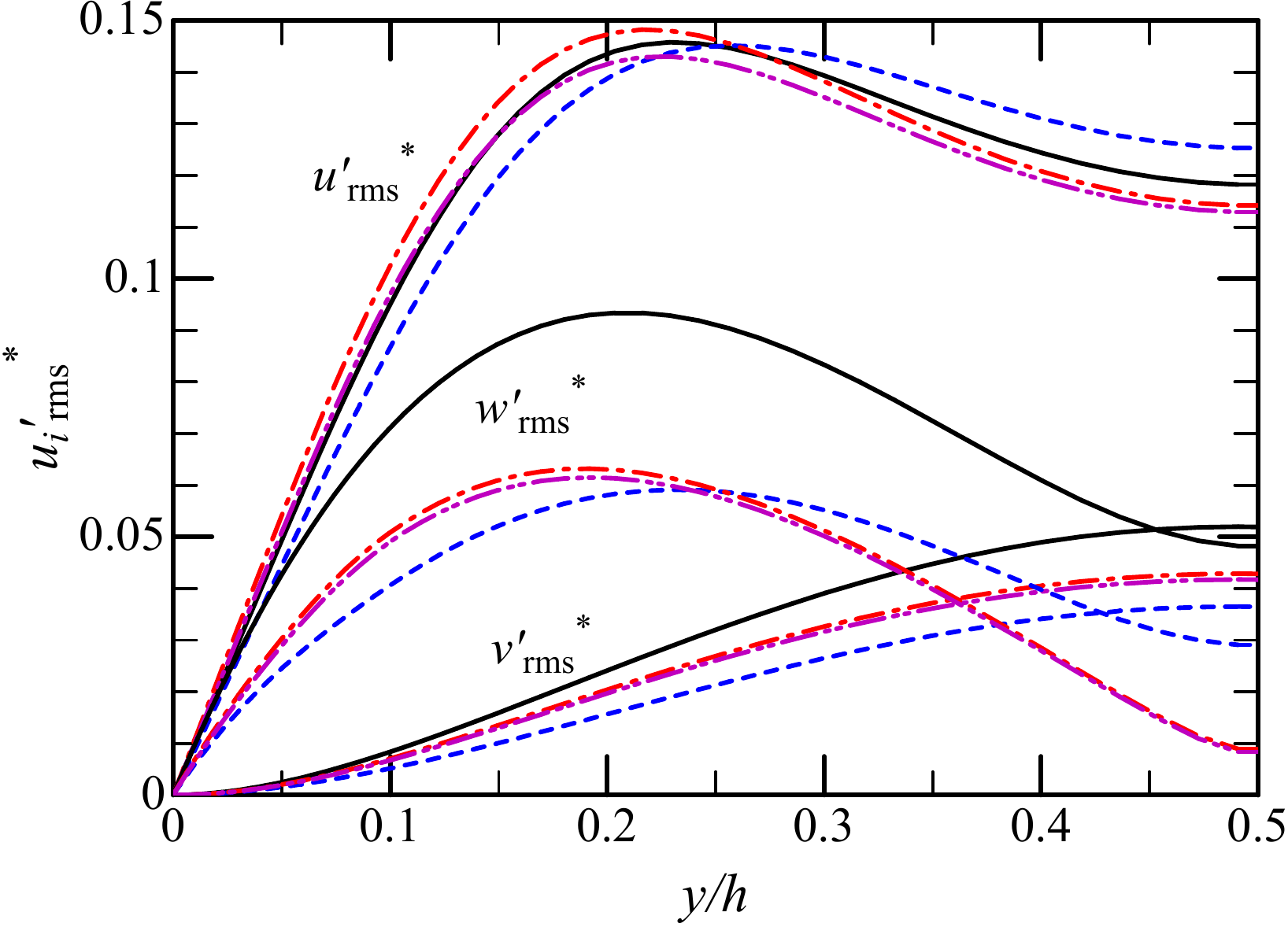}
 \caption{Wall-normal profiles of the velocity fluctuation intensity normalized by $U_\mathrm{w}$. The legend is the same as in \Fref{fig:mean_u}.}
 \label{fig:rms}
\end{center}
\end{figure}%

\Fref{fig:rms} presents the profiles of the root-mean-square of the velocity fluctuations normalized by $U_\mathrm{w}$. 
The streamwise velocity fluctuation $u^\prime$ is relatively unaffected by the addition of the viscoelasticity, and the wall-normal and spanwise components are significantly reduced in the $\Wew=1000$ case compared to the Newtonian cases. 
In particular, the reduction in the spanwise component $w^\prime_\mathrm{rms}$ is remarkable, and  the peak level of $w^\prime_\mathrm{rms}$ in the case of $\Wew=1000$ is reduced by about 34\% from the Newtonian case. 
Such a significant suppression of $w_\mathrm{rms}$ is consistent with the observation in the instantaneous flow field shown in figures~\ref{fig:zy_u}(a) and (b). 
As for the higher $\Wew$ cases, the Weissenberg number effect in the higher $\Wew$ are not so remarkable, and in particular the results of $\Wew=1500$ and 2000 cases are almost identical. 

\begin{figure}
\begin{center}
 \includegraphics[height=70mm]{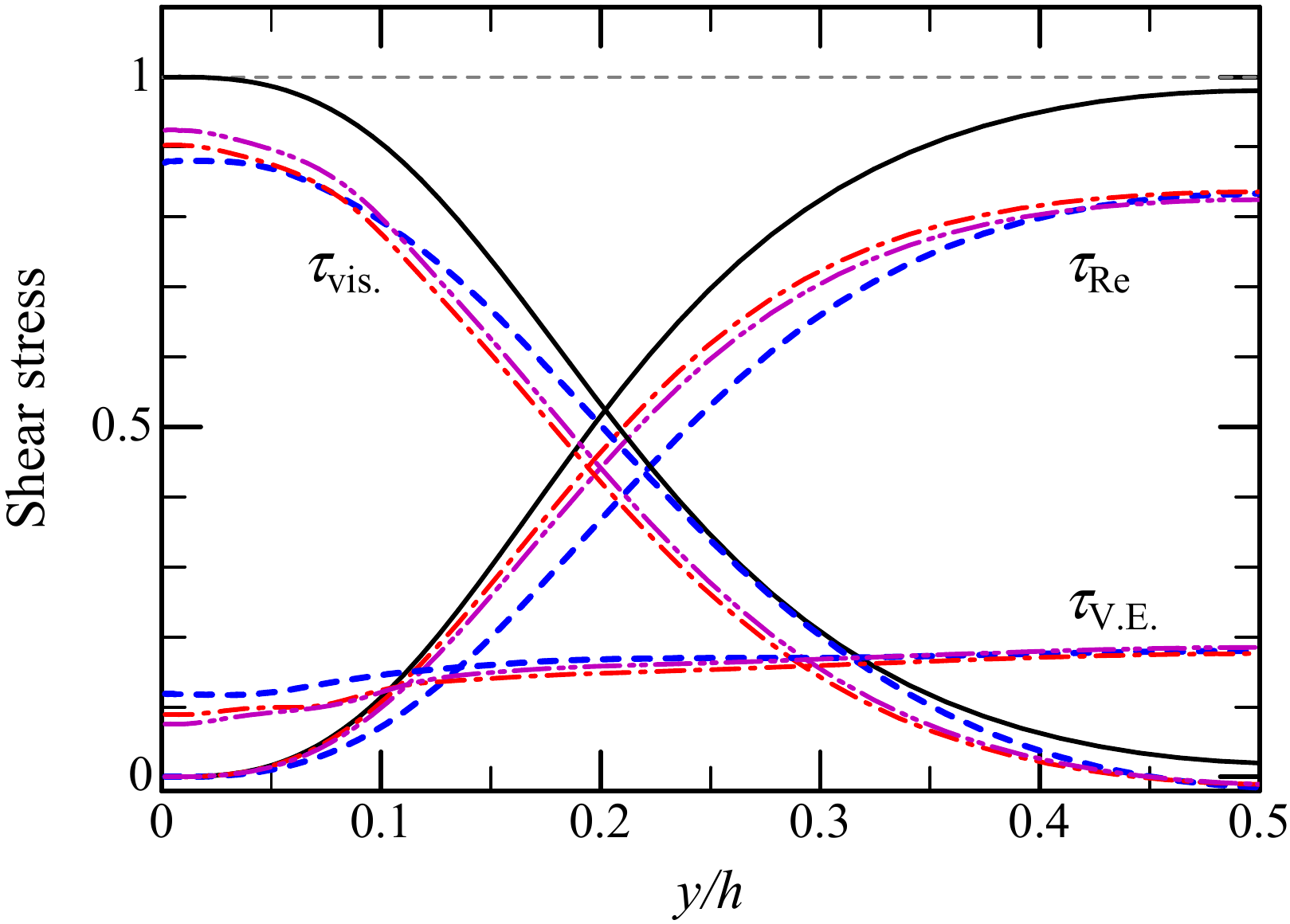}
 \caption{Shear stress balance scaled by the wall-shear stress: the viscous $\tau_{\rm vis.}$, Reynolds $\tau_{\rm Re}$, and viscoelastic shear stress $\tau_{\rm V.E.}$. In the plane Couette flow, they should satisfy $\tau_{\rm vis.} + \tau_{\rm Re} + \tau_{\rm V.E.} = 1$, when normalized by the wall shear stress. The legend is the same as in \Fref{fig:mean_u}.}
 \label{fig:stress}
\end{center}
\end{figure}%

The shear stresses presented in \Fref{fig:stress} are non-dimensionalized by $\tau_\mathrm{w}$ in each case. 
In the Newtonian case, the Reynolds shear stress $\tau_{\rm Re}=-\overline{u^\prime v^\prime}$ accounts for most of the total shear stress at the channel center, and the sum of $\tau_{\rm Re}$ and the viscous shear stress $\tau_{\rm visc.}$ are equal to 1 across the channel. 
On the other hand, in the viscoelastic cases, the viscous and Reynolds stress contributions to the total shear stress decrease by the addition of the viscoelastic shear stress $\tau_{\rm V.E.} = (1-\beta)\overline{c_{12}}/\Wew$. 
It is also seen that the profiles of the three shear stresses for the different viscoelastic cases relatively well collapse despite of the difference in the Weissenberg number. 
The viscoelastic shear stress $\tau_{\rm V.E.}$ is particularly less affected by $\Wew$, and the variation across the channel is also rather constant. 
Comparing the Reynolds shear stress contribution for different $\Wew$ cases, one can see that $\tau_{\rm Re}$ at the channel center is hardly affected by $\Wew$ and accounts for about 80\% of the total shear stress, which is interestingly consistent with the relative concentration of the Newtonian fluid ($\beta=0.8$). 

\subsection{Reynolds stress transport}
To further discuss the energy exchange between the flow filed and additive in the viscoelastic flow, the budget of the Reynolds-stress transport equations is investigated.  
In the viscoelastic case, the Reynolds-stress transport equation is described as:
\begin{equation} 
\left( \frac{\partial}{\partial t^\ast} + \overline{u_k^\ast} \frac{\partial}{\partial x_k^\ast} \right) \overline{{u_i^\prime} {u_j^\prime}}^\ast = P_{ij} + \Pi_{ij} + D_{ij} - \varepsilon_{ij}  + G_{ij} +W^{ve}_{ij}, \label{eq:rss}
\end{equation}
where $P_{ij}$, $\Pi_{ij}$, and $\varepsilon_{ij}$ are the production, pressure-strain redistribution, and viscous dissipation terms, $D_{ij}$ stands for the sum of the viscous, pressure, and turbulent transport terms, and $G_{ij}$ represents the effect of the Coriolis force:
\begin{equation}
G_{ij}=-\frac{\Omega}{\Rew} \left( \epsilon_{i3k} \overline{{u_k^\prime} {u_j^\prime}}^\ast + \epsilon_{j3k} \overline{{u_k^\prime} {u_i^\prime}}^\ast \right). 
\label{eq:gij}
\end{equation}
The last term in the right-hand-side of \Eref{eq:rss} is the viscoelastic-force term: 
\begin{eqnarray}
W^{ve}_{ij}=\overline{{u_i^\prime}^\ast {f^{ve}_j}^\prime} + \overline{{u_j^\prime}^\ast {f^{ve}_i}^\prime} \quad
\textrm{where, } {f^{ve}_i}^\prime = \frac{1-\beta}{\Wew} \frac{\partial c_{ik}^\prime}{\partial x_k^\ast} \nonumber.
\end{eqnarray}
In particular for the fluctuating kinetic energy $k=\overline{{u_i^\prime} {u_i^\prime}}^\ast /2$, $W^{ve}_{ij}$ is the averaged inner product between the fluctuating velocity and viscoelastic-force vectors, which represents the averaged work done by the fluctuating viscoelastic force to the fluctuating flow field. 

%%%%%%% Figure 9 %%%%%%%%%
\begin{figure}[t]
\begin{center}
	\begin{minipage}{0.45\hsize}
		\includegraphics[width=1\hsize]{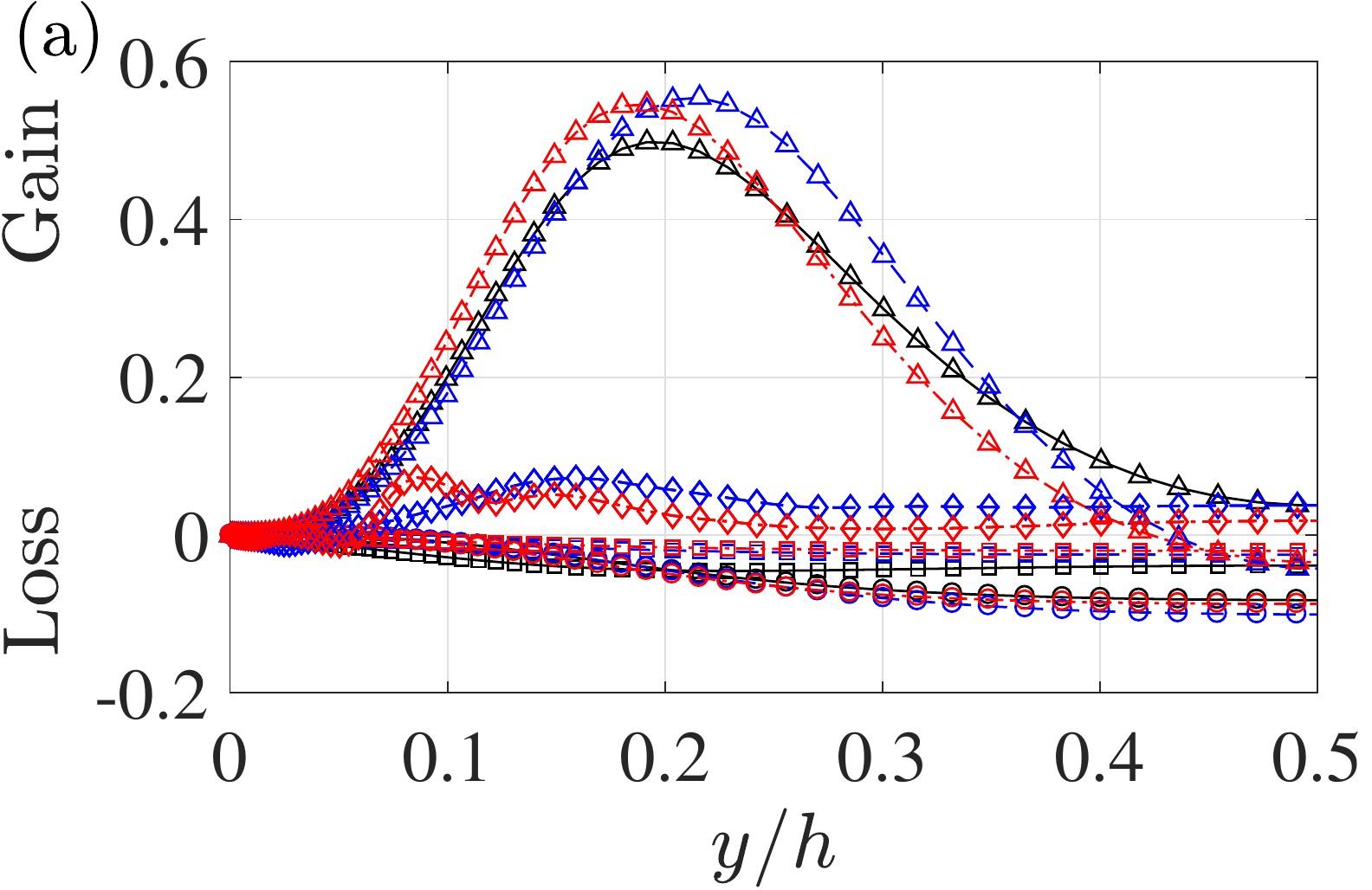}
	\end{minipage} \hspace{1em}
	\begin{minipage}{0.45\hsize}
		\includegraphics[width=1\hsize]{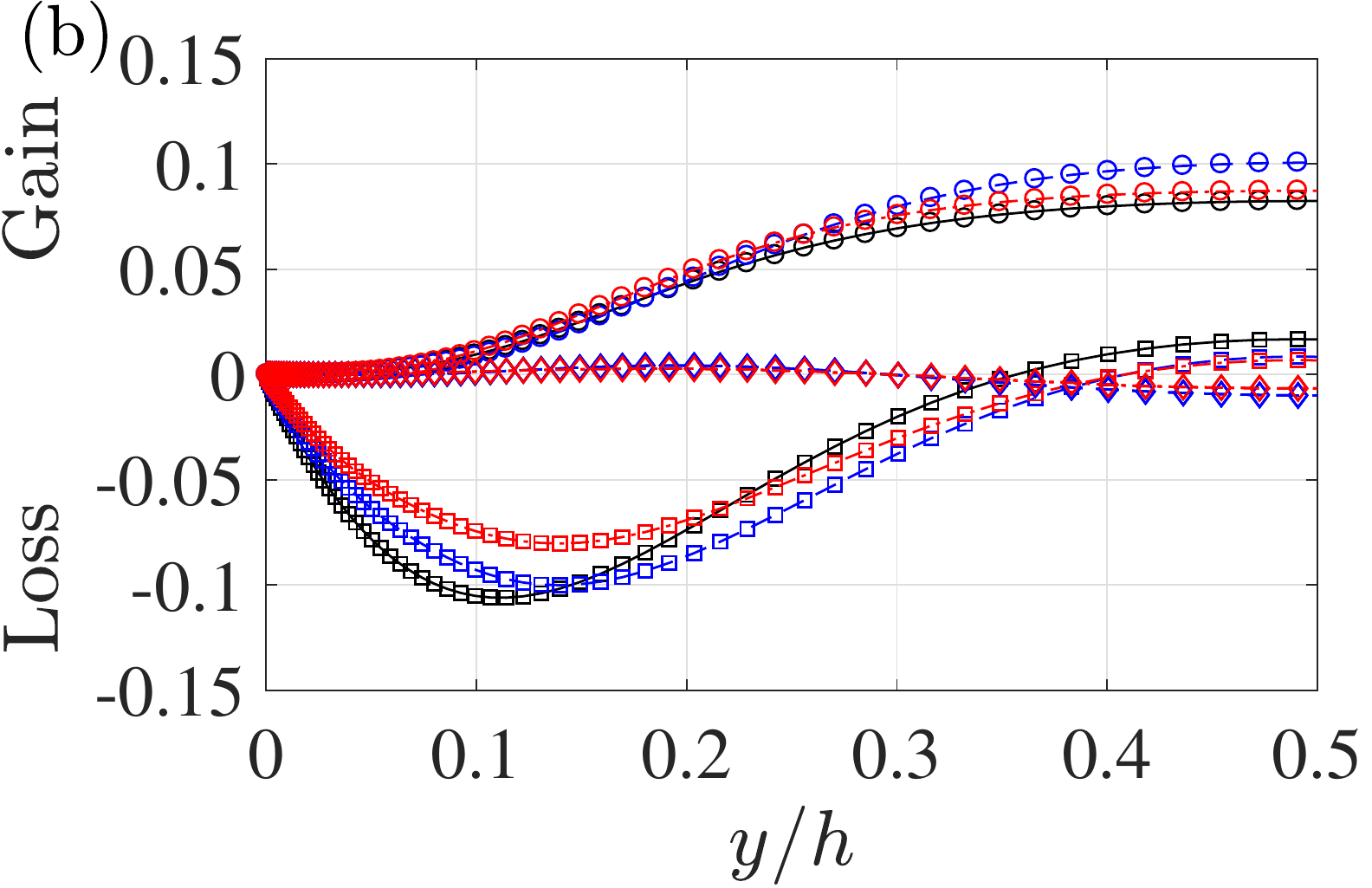}
	\end{minipage} \\
	\begin{minipage}{0.45\hsize}
		\includegraphics[width=1\hsize]{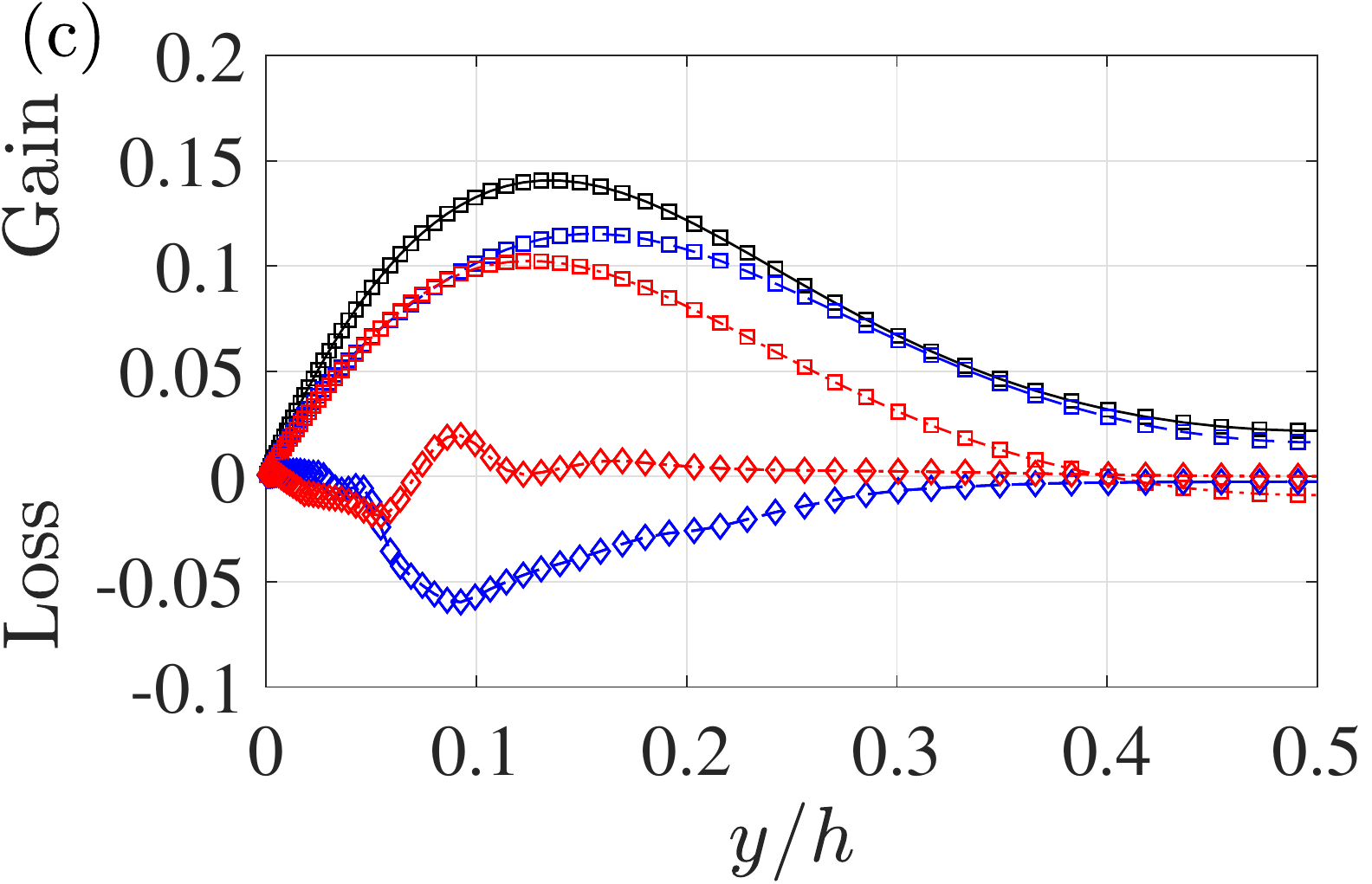}
	\end{minipage} \hspace{1em}
	\begin{minipage}{0.45\hsize}
		\includegraphics[width=1\hsize]{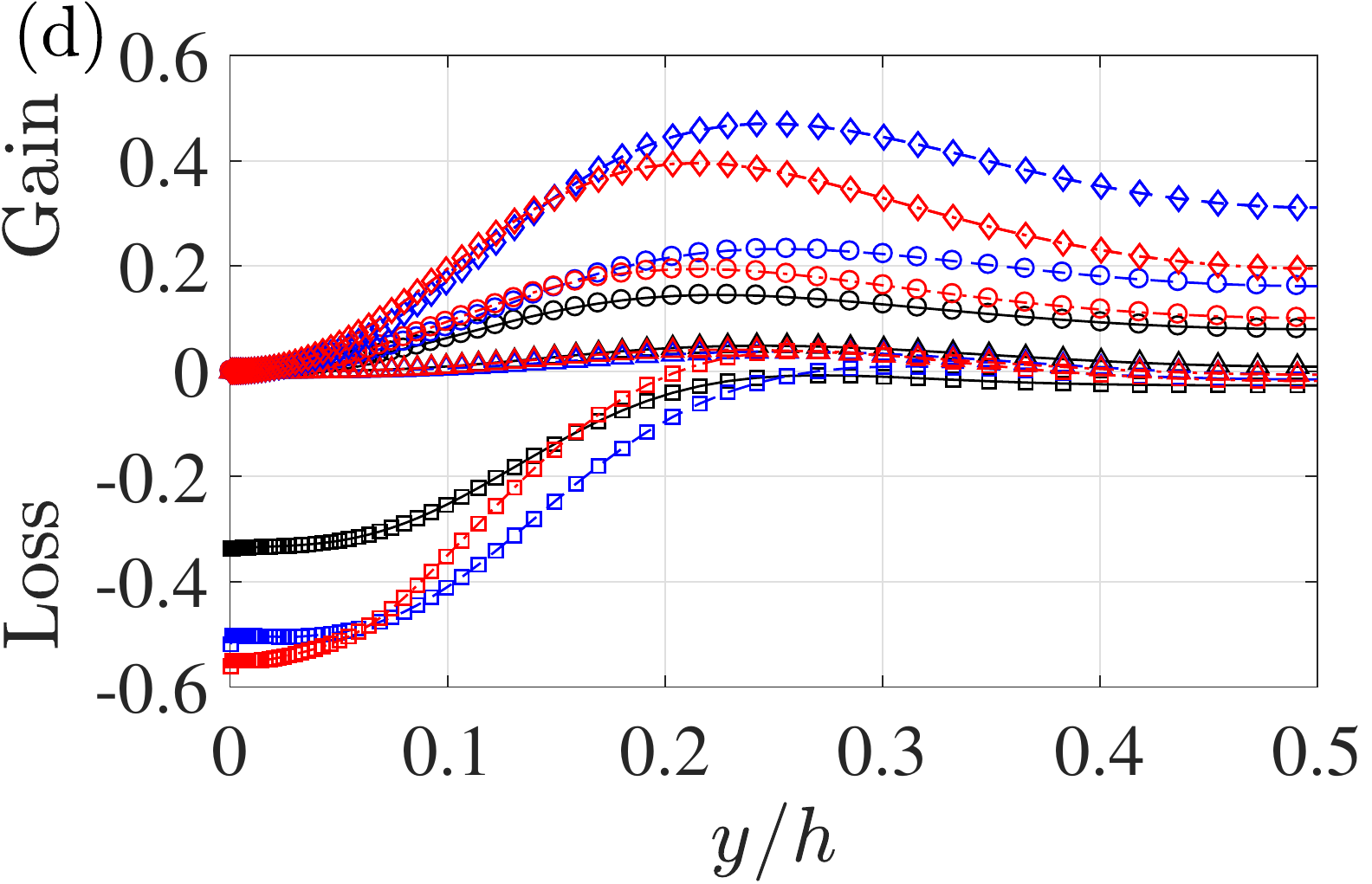}
	\end{minipage}
\caption{Budget of transport equations of the Reynolds stresses (a) $\overline{{u^\prime}^2}$, (b) $\overline{{v^\prime}^2}$, (c) $\overline{{w^\prime}^2}$, and (d) $-\overline{u^\prime v^\prime}$. The values are scaled by $u_\tau^4/\nu$. The colors of the lines in each panel represent (black) the Newtonian case, (blue) $\Wew=1000$ case, and (red) $\Wew=1500$ case, and the symbols represent the different terms in the Reynolds stress transport equation: 
\opentriangle, production term $P_{ij}$; 
\opencircle, Coriolis-force term $G_{ij}$; 
\opensquare, pressure-strain redistribution term $\Pi_{ij}$;
\opendiamond, viscoelastic-force term $W_{ij}$. }
\label{fig:budget}
\end{center}
\end{figure}
% %%%%%%%%% Figure 9 %%%%%%%%%%%%%%%

\Fref{fig:budget} presents the profiles of $P_{ij}$, $G_{ij}$, $\Pi_{ij}$, and $W^{ve}_{ij}$ of the transport equations of the Reynolds normal stresses $\overline{{u^\prime}^2}$, $\overline{{v^\prime}^2}$, $\overline{{w^\prime}^2}$, and the Reynolds shear stress $-\overline{u^\prime v^\prime}$, compared for the Newtonian, $\Wew=1000$, and $\Wew=1500$ cases. The values shown in the figure are scaled based on the contribution from the Newtonian solvent to the friction velocity: that is $u_\tau^\ast=\sqrt{\beta \Psi^\ast |_{y=0} /\Rew}$. 
The case of $\Wew=2000$ is not included here, because the results are almost identical to those for $\Wew=1500$.  

As for the energy source, for each Reynolds normal-stress component, only the streamwise component $\overline{{u^\prime}^2}$ has the non-zero energy input from the mean flow gradient through the term of
\begin{equation}
P_{11}^\ast=-2 \overline{u^\prime v^\prime}^\ast \frac{\mathrm{d} \overline{u}^\ast}{\mathrm{d} y^\ast},
\end{equation}
as shown in \Fref{fig:budget}(a), and $P_{11}$ is the significant source of the $\overline{{u^\prime}^2}$ component for both fluids. 
The main energy source for the wall-normal component $\overline{{v^\prime}^2}$ is the Coriolis-force term $G_{22}$, as shown in \Fref{fig:budget}(b). 
From \Eref{eq:gij}, one can easily see that $G_{11}$ and $G_{22}$ are: 
\begin{eqnarray}
G_{11}^\ast=2\frac{\Omega}{\Rew} \overline{ u^\prime v^\prime}^\ast 
\quad {\rm and} \quad
G_{22}^\ast=-2\frac{\Omega}{\Rew} \overline{ u^\prime v^\prime}^\ast, \nonumber
\end{eqnarray}
respectively, indicating that the Coriolis-force term $G_{ij}$ is the inter-component energy transfer from the $\overline{{u^\prime}^2}$ to $\overline{{v^\prime}^2}$ component. 
For the spanwise normal component $\overline{{w^\prime}^2}$, both the production term $P_{33}$ and the Coriolis energy-transfer term $G_{33}$ are zero, and the source for this component is the pressure-strain redistribution term $\Pi_{33}$, as shown in \Fref{fig:budget}(c). 
As the pressure strain terms for $\overline{{u^\prime}^2}$ and $\overline{{v^\prime}^2}$ components are negative (see \Fref{fig:budget}(a) and (b)), this $\Pi_{ij}$ term redistributes the energy from the $\overline{{u^\prime}^2}$ and $\overline{{v^\prime}^2}$ components to the $\overline{{w^\prime}^2}$. 
Therefore, only the $\overline{{u^\prime}^2}$ gains energy from the mean flow field, and the other lateral normal components are fed by an inter-component energy transfer by either the Coriolis-force term or the pressure-strain redistribution term. 
This tendency of the energy transport is basically the same for all the present Newtonian/viscoelastic cases. 
A difference with the non-rotating near-wall turbulence at high Reynolds number is the main energy source of $\overline{{v^\prime}^2}$.
In the non-rotating case, $\Pi_{22}$ should be a dominant energy gain, since the Coriolis term of $G_{22}$ is absent. 
There seems to be a positive-$\Pi_{22}$ region in the channel central region even in the present RPCF.
Despite this difference in the $\overline{{v^\prime}^2}$ energy source, the net energy transported among the respective components may be regarded as similar between the near-wall turbulence and the present target in RPCF.

Focusing on the role of the viscoelastic-force term $W^{ve}_{ij}$ in the Reynolds stress transport, one can see in \Fref{fig:budget}(a) that $W^{ve}_{ij}$ behaves as a source for $\overline{{u^\prime}^2}$, almost compensating for the energy loss by the Coriolis force term $G_{11}$. The magnitude of $W^{ve}_{22}$ is almost zero throughout the channel. 
This tendency is basically the same for different Weissenberg number cases. 
The spanwise component $W^{ve}_{33}$ is relatively smaller than $W^{ve}_{11}$ in magnitude, as shown in \Fref{fig:budget}(c), and it works as a sink term for $\overline{{w^\prime}^2}$ at $\Wew=1000$. 
In the case of $\Wew=1500$, the wall-normal profile of $W^{ve}_{33}$ (which is spatially averaged) reveals a rather complicated distribution around zero. 
This term might play a key role in locally suppressing the spanwise secondary flow that induces the wavy motion of roll cells, as discussed later.
Therefore, the viscoelastic-force mainly affects the streamwise and spanwise velocity fluctuations. 

As described in the previous section the secondary motion of the roll cells in the viscoelastic cases are relatively suppressed compared to the Newtonian case, while the magnitude of the high and low-speed streaks are maintained. 
Such a tendency can be explained by the above-mentioned aspects of the viscoelastic-force term ($W^{ve}_{11}$ and $W^{ve}_{33}$), which tends to support the streamwise velocity fluctuation while reducing the spanwise component in the diagonal spanwise secondary flow observed in the 3D wavy roll cells. 
The suppression in secondary flow motion (as well as the wavy pattern) in the viscoelastic flows is also attributable to the change in the pressure-strain terms of $\Pi_{11}$ and $\Pi_{33}$. 
As $\Wew$ increases, $\Pi_{11}$ is suppressed in the viscoelastic cases and, corresponding to this, $P_{33}$ also decreases. 
This is because the roll-cell structures at the $\Wew=1500$ and 2000 cases are streamwise-independent with almost zero $\partial u^\prime/\partial x$, and thereby the energy redistribution from the streamwise to spanwise velocity component,
\begin{equation}
\Pi_{11} = 2 \overline{p^\prime \frac{\partial u^\prime}{\partial x}}^\ast
\end{equation}
is suppressed. 
Therefore, the modulation in `shape' of roll cells can inhibit the secondary flow motion of the structure through the pressure-strain redistribution. 

As for the gain-and-loss balance in the equation of the Reynolds shear stress $-\overline{u^\prime v^\prime}$, \Fref{fig:budget}(d) shows that the Coriolis-force term $G_{-12}$ and the pressure-strain term $P_{-12}$ are dominant as the source and sink, respectively. The Coriolis-force term $G_{-12}$ and the production by the mean velocity gradient for the $-\overline{u^\prime v^\prime}$ component are
 \begin{equation}
G_{-12}^\ast=\frac{\Omega}{\Rew} \left( \overline{{u^\prime}^2}^\ast - \overline{{v^\prime}^2}^\ast \right)
\quad {\rm and} \quad 
P_{-12}=\overline{{v^\prime}^2}^\ast \frac{\mathrm{d} \overline{u}^\ast}{\mathrm{d} y^\ast}, \label{eq:-12}
 \end{equation}
respectively. 
Although the production by the mean flow is not zero for the shear stress component, $P_{-12}$ is not comparable to $G_{-12 }$ as shown in \Fref{fig:budget}(d). 
Irrespective of the kinds of fluid, the shear stress $-\overline{u^\prime v^\prime}$ is mainly produced by the Coriolis-force term in the channel central region, carried towards the near-wall region by the transport terms (not shown in figure), and dissipated by $P_{-12}$. 
It is interesting to note that the viscoelastic term of $W^{ve}_{-12}$ works as a significant positive source of the Reynolds shear stress throughout the channel. 
As discussed in this paper, the turbulent intensities in the directions normal to the main stream are suppressed by the elasticity, in particular, $v'_{\rm rms}$ is damped well at $\Wew=1000$ (see \Fref{fig:rms}). 
On the contrary, a direct effect of the elasticity via $W^{ve}_{-12}$ is largest at $\Wew=1000$, which would compensate the damping of the $v'$ magnitude. 
This numerical fact may explain the non-significant decrease in $\tau_\mathrm{Re}$ as observed in \Fref{fig:stress}, and provide an indication to elucidate the longstanding issue of a difference between the non-zero and almost-zero Reynolds shear stresses in highly drag-reducing flows demonstrated respectively by DNS and by experimentation \cite{Yu04b,tsuka11}. 
As shown by \Eref{eq:-12} the Coriolis-force term $G_{-12}$ is proportional to the anisotropy between $\overline{{u^\prime}^2}$ and $\overline{{v^\prime}^2}$, and the viscoelastic-force term $W^{ve}_{11}$ and $W^{ve}_{22}$ enhances the difference between them as mentioned previously. 
Therefore, the elasticity also affects the production of the Reynolds shear stress $-\overline{u^\prime v^\prime}$ indirectly by promoting the anisotropy between the streamwise and wall-normal velocity fluctuations. 

%%%%%%% New Section %%%%%%%%%%%%%%%%%%%%%%%%%%%%%%%%%%%%%%%%%%%%%%%%%%%%%%%%%%%%
\section{Discussion}

In the previous section, the addition of elasticity was shown to modulate the shape of the roll-cell structure, transforming the streamwise-dependent wavy roll cells observed in the Newtonian case into the 2D streamwise-independent roll cells. 
It was also shown that the secondary flow motion in the roll cells affected by elasticity is suppressed compared to the 3D roll cells in the Newtonian case, while the roll-cell induced streamwise velocity fluctuation was not significantly affected. 
The investigation into the Reynolds-stress transport equations indicates that the viscoelastic-force term plays a significant role in the transport equations by supporting the streamwise velocity fluctuation while suppressing the spanwise component. 

%%%%%%%% Figure 10 %%%%%%%%%%
\begin{figure}
\begin{center}
 \includegraphics[width=0.7\hsize]{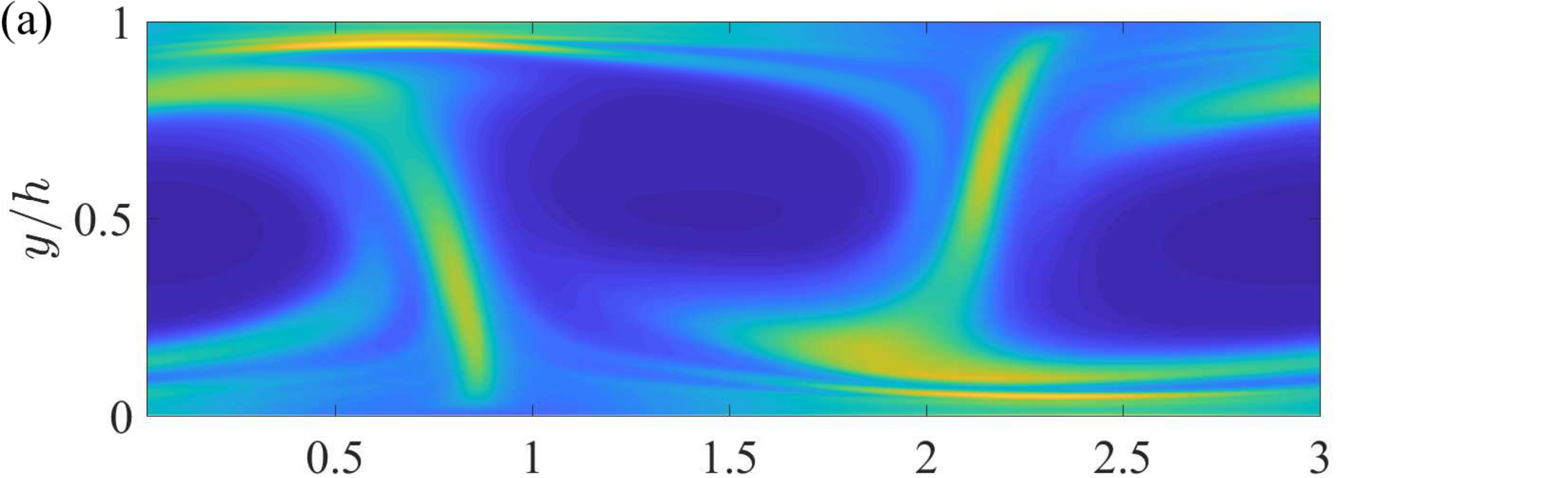}
 \includegraphics[width=0.7\hsize]{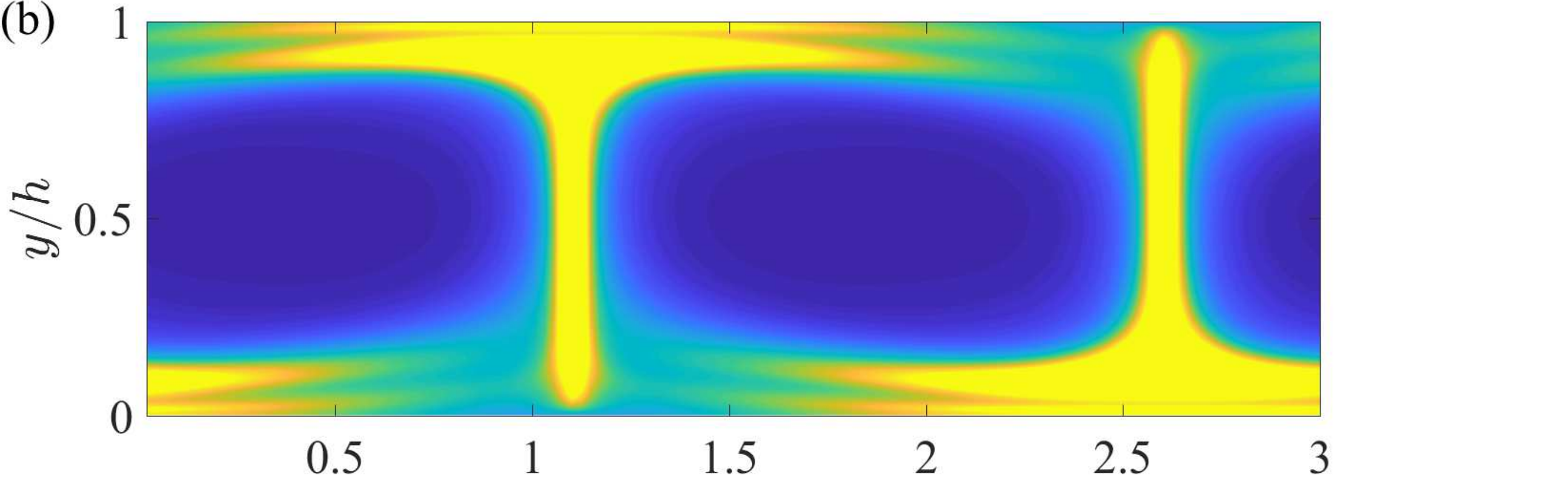}
 \includegraphics[width=0.7\hsize]{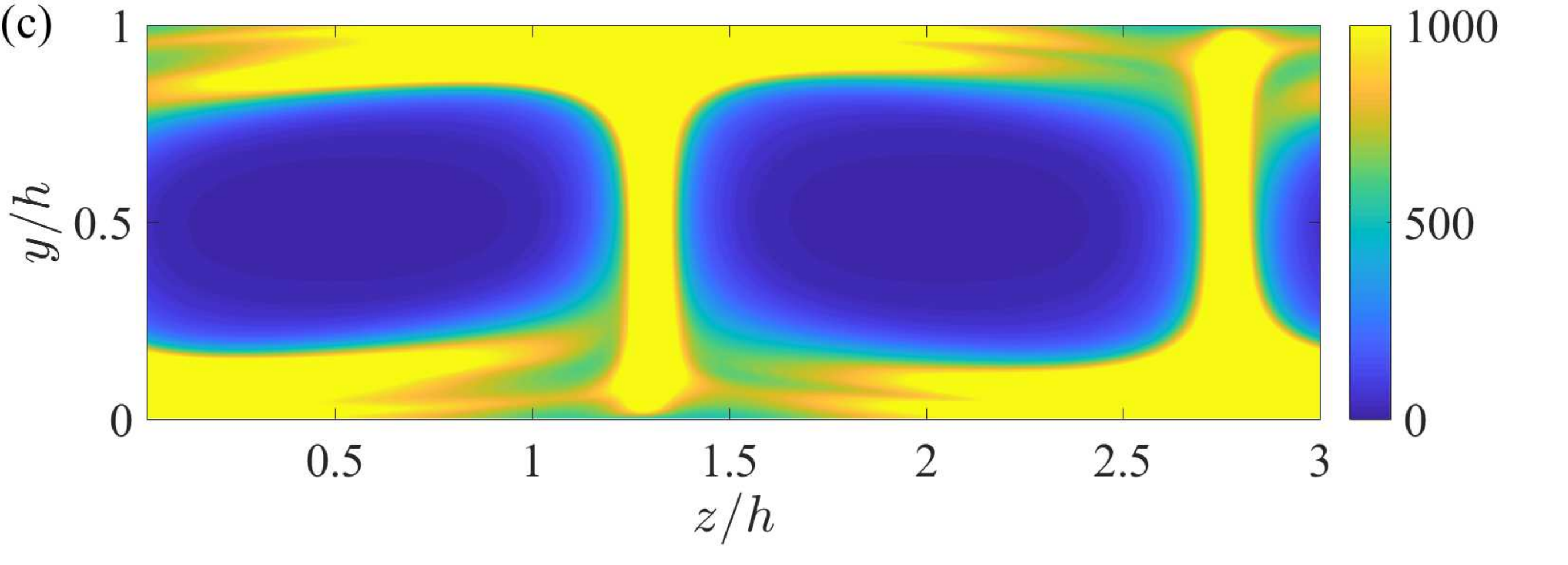}
 \caption{Distribution of the trace of the viscoelastic stress tensor $c_{11}+c_{22}+c_{33}$ on the cross-sectional plane at the same streamwise positions as in \Fref{fig:zy_u} for the case of (a) $\Wew=1000$, (b) $\Wew=1500$, and (c) $\Wew=2000$. }
 \label{fig:zy_ve}
\end{center}
\end{figure}
%%%%%%%%%% Figure 10 %%%%%%%%%%%

By averaging \Eref{eq:Giesekus}, one obtains the transport equation of the average viscoelastic stress tensor $\overline{c_{ij}}$:
\begin{eqnarray}
\left( \frac{\partial}{\partial t^\ast} + \overline{u_k^\ast} \frac{\partial}{\partial x_k^\ast} \right) \overline{c_{ij}} =&
\underbrace{\overline{c_{ik}} \frac{\partial \overline{u_j}^\ast}{\partial x_k^\ast} + \overline{c_{jk}} \frac{\partial \overline{u_i}^\ast}{\partial x_k^\ast}}_{S^{VE}_{ij}}
+\underbrace{\overline{c_{ik}^\prime \frac{\partial {u_j^\prime}^\ast}{\partial x_k^\ast}}+\overline{c_{jk}^\prime \frac{{\partial u_i^\prime}^\ast}{\partial x_k^\ast}}}_{S^{ve}_{ij}} 
-\frac{\partial  \overline{{u_k^\prime}^\ast c_{ij}^\prime} }{\partial x_k^\ast} \nonumber \\
&-\frac{\Rew}{\Wew} \overline{ \left(c_{ij}-\delta_{ij} + \alpha (c_{im}-\delta_{im})(c_{mj}-\delta_{mj}) \right)},
\label{eq:cij}
\end{eqnarray}
where the first and second terms in the right-hand side indicate the deformation work done by the mean and fluctuating flow field to the additive, respectively, and the other terms are the spatial transport terms of the flow field and the averaged Giesekus model term. 
The deformation work $S^{VE}_{ij}$ and $S^{ve}_{ij}$, through which the viscoelastic stress $c_{ij}$ gains the energy from the flow fields, can be related to the viscoelastic-force term $W^{VE}_{ij}$ and $W^{ve}_{ij}$ in the Reynolds-stress transport equation, \Eref{eq:rss} as
\begin{eqnarray}
S^{VE}_{ij} &= -\frac{\Wew}{1-\beta} {W^{VE}_{ij}}^\ast + \frac{\partial }{\partial x_k^\ast} \left( \overline{u_i}^\ast \: \overline{c_{kj}} + \overline{u_j}^\ast \: \overline{c_{ki}}\right), \label{eq:SEV}\\
S^{ve}_{ij} &=-\frac{\Wew}{1-\beta} {W^{ve}_{ij}}^\ast + \frac{\partial }{\partial x_k^\ast} \left( \overline{{u_i^\prime}^\ast c_{kj}} + \overline{{u_j^\prime}^\ast c_{ki}}\right),\label{eq:Sev}
\end{eqnarray}
where $W^{VE}_{ij}=\overline{u_i} \; \overline{f^{ve}_j} + \overline{u_j} \; \overline{f^{ve}_i}$ is the mean viscoelastic-force term (here, $f^{ve}_i$ is the viscoelastic force in the $i$-direction). 
As the equations above show, $S^{VE}_{ij}$ and $W^{VE}_{ij}$ (also between $S^{ve}_{ij}$ and $W^{ve}_{ij}$) are connected by each additional spatial transport term, indicating that their total amounts integrated across the channel are always equivalent to each other. 

\Fref{fig:zy_ve} presents the instantaneous distributions of the trace of the viscoelastic stress tensor ${\rm tr}(c_{ij})$, which indicates the degree of the total stretching of additive (polymer). 
The value of ${\rm tr}(c_{ij})$ is almost zero inside the vortex while it is significantly enlarged near the edge of the roll cells, particularly in the near wall regions where the in-plane flow direction is parallel to the wall. 
The magnitude of ${\rm tr}(c_{ij})$ increases with the increasing Weissenberg number.

%%%%%%% Figure 12 %%%%%%%%%%%%%%%
\begin{figure}
\begin{center}
 \includegraphics[height=70mm]{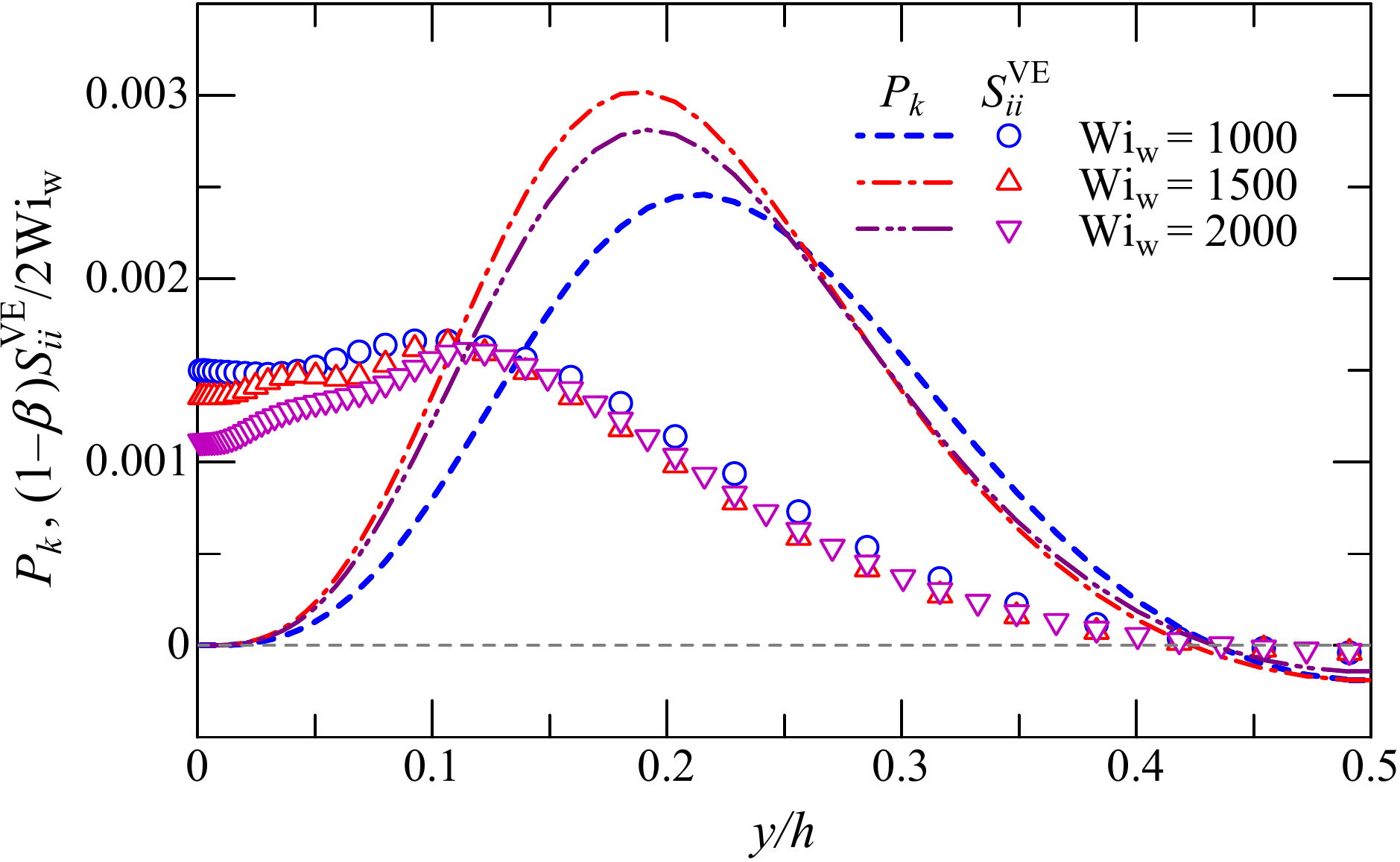}
 \caption{Mean flow production of the averaged viscoelastic stress $\overline{c_{ii}}$ and the turbulent kinetic energy production $P_k$. The values are scaled by the wall speed $U_\mathrm{w}$ and the channel gap $h$.}
 \label{fig:SVE11}
\end{center}
\end{figure}
%%%%%%% Figure 12 %%%%%%%%%%%%%%%

Among the normal stresses, only the streamwise component $\overline{c_{11}}$ has a non-zero production by the mean shear flow. 
Hence, the production of the trace of $c_{ij}$ by the mean flow is:
\begin{equation}
S^\mathit{VE}_{ii}=S^\mathit{VE}_{11}=\overline{c_{12}}^\ast \frac{\mathrm{d} \overline{u}^\ast}{\mathrm{d} y^\ast}.
\end{equation} 
As the other normal stress components do not have a mean-shear production, the streamwise component $\overline{c_{11}}$ dominates the other components. 
\Fref{fig:SVE11} compares the production of the trace of the viscoelastic stress tensor $S^\mathit{VE}_{ii}$ to the turbulent kinetic energy production $P_{k}$ for different Weissenberg number cases. 
In the figure, $S^\mathit{VE}_{ii}$ is multiplied by $(1-\beta)/2 \Wew$ so that $S^\mathit{VE}_{ii}$ and $P_k$ are fairly comparable in terms of the kinetic energy loss in the mean flow. 
It shows that the mean flow provides comparable amount of energy to the viscoelastic stress as it does to the fluctuating flow field. 
The integrated value of $(1-\beta)S^\mathit{VE}_{ii}/2\Wew$ across the channel is 79\% of that of $P_k$ in the case of $\Wew=1000$, and about 60\% in the higher $\Wew$ cases. 
It is also shown that $S^\mathit{VE}_{ii}$ does not decrease in the near wall region unlike $P_k$, as $\overline{c_{12}}$ does not decrease in the vicinity of the wall as shown in \Fref{fig:rms}(b). 
Owing to the production by the mean shear flow, $c_{11}$ is significant in the near wall region as shown in \Fref{fig:zy_ve}. 
A part of the significant amount of energy from the mean flow to $\overline{c_{11}}$ is also transferred to the Reynolds normal stress component $\overline{{u^\prime}^2}$ by $S^{ve}_{11}$ (i.e., $W^{ve}_{11}$), as shown in \Fref{fig:budget}(a). 

%%%%%%%%%%% Figure 13 %%%%%%%%%%%%%%%
\begin{figure}[t]
\begin{center}
 \includegraphics[width=0.7\hsize]{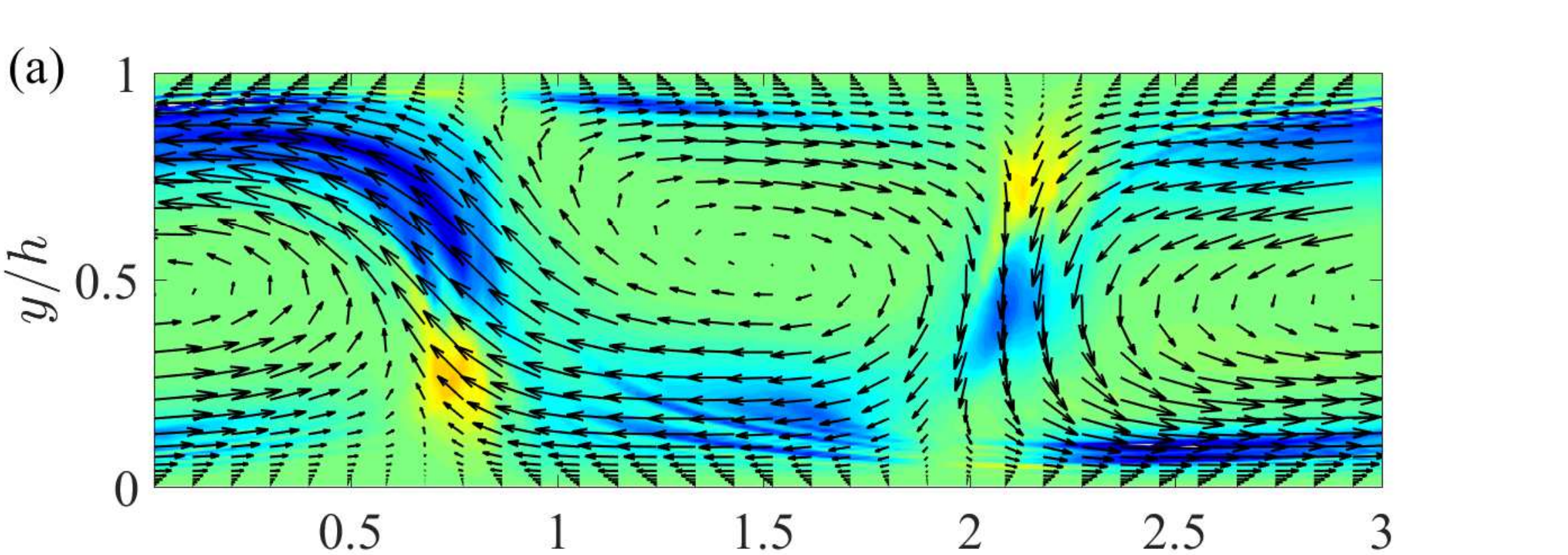}\\
 \includegraphics[width=0.7\hsize]{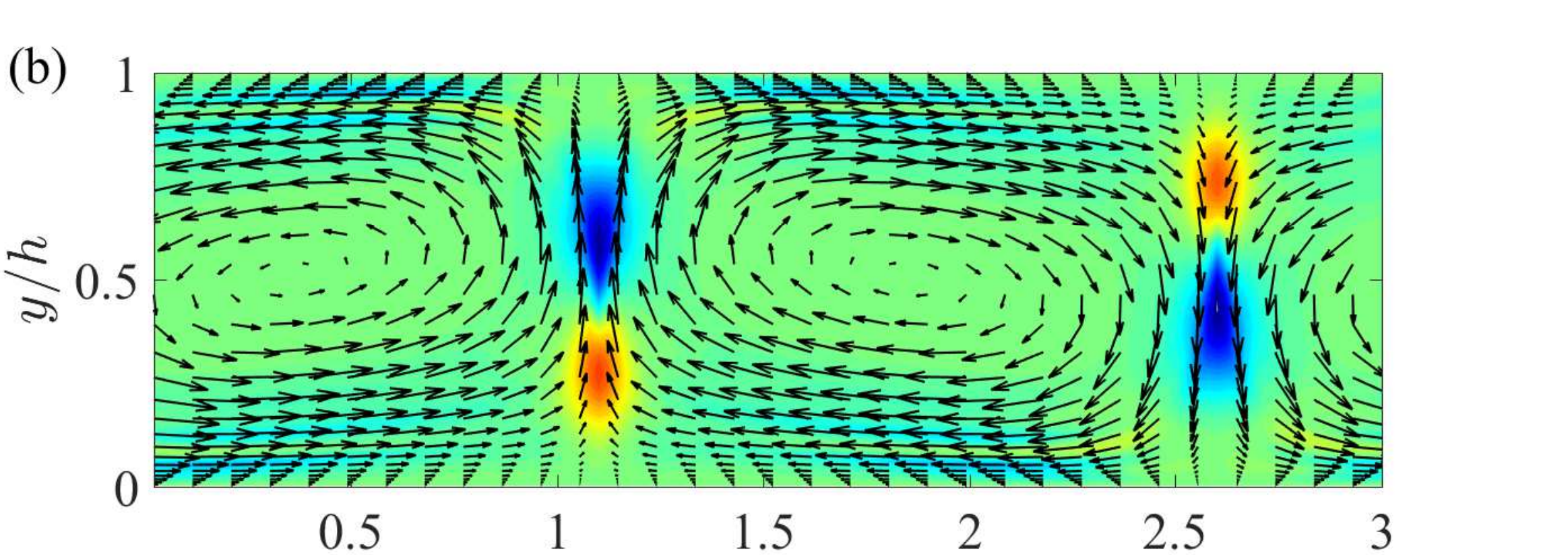}
 \includegraphics[width=0.7\hsize]{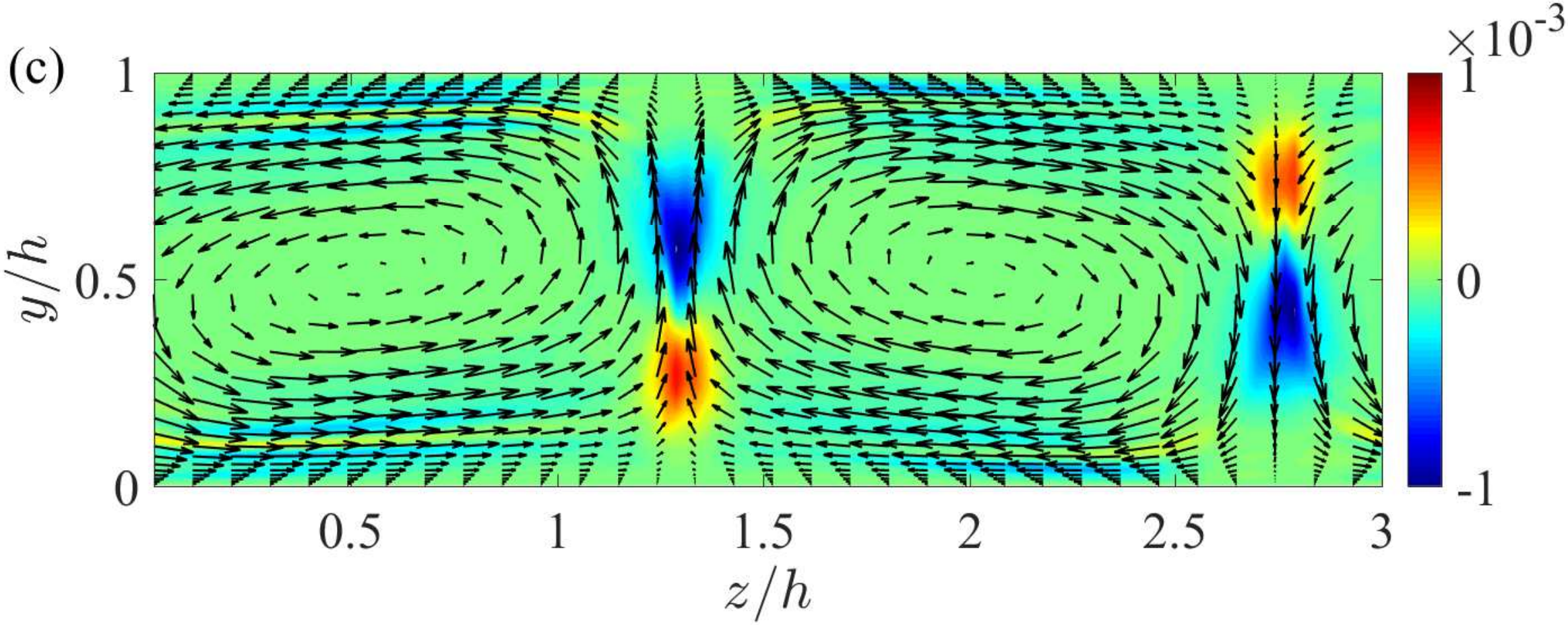}
 \caption{Distribution of $v^\prime {f^{ve}_y}^\prime + w^\prime {f^{ve}_z}^\prime$ on the cross-sectional planes at the streamwise positions indicated by the white dashed line in \Fref{fig:xz_v}: (a) $\Wew=1000$, (b) $1500$, and (c) $2000$. 
The values indicated by the colors are normalized by $U_\mathrm{w}$ and $h$, and the black arrows represent the in-plane velocity vector pattern $(v^\prime, w^\prime)$.}
 \label{fig:workzy}
\end{center}
\end{figure}
%%%%%%%%%%% Figure 13 %%%%%%%%%%%%%%%

Conversely, $\overline{c_{22}}$ and $\overline{c_{33}}$ do not have the production of the mean flow unlike $\overline{c_{11}}$, and the energy exchange between the fluctuating fields of velocity and conformation tensor is not significant for these (wall-normal and spanwise) components, as shown in \Fref{fig:budget}; however, this is not true for the $\Wew=1000$ case, in which $\overline{c_{33}}$ receives some energy from $\overline{{w^\prime}^2}$. 
\Fref{fig:workzy} presents the cross-sectional distribution of the inner-product between the in-plane velocity vector $(v^\prime, w^\prime)$ and viscoelastic-force vector $({f^{ve}_y}^\prime, {f^{ve}_z}^\prime)$, which indicates the energy exchange between ${v^\prime}^2+{w^\prime}^2$ and $c_{22}+c_{33}$. 
The average of this quantity in the $x$- and $z$-directions and in time yields in $W^{ve}_{22}+W^{ve}_{33}$. 
At $\Wew=1000$, $v^\prime {f^{ve}_y}^\prime+w^\prime {f^{ve}_z}^\prime$ is significantly negative on the edge of the roll cells in the near wall region, where the secondary flow is mainly in the spanwise direction, indicating that in such regions the in-plane viscoelastic-force resists against the spanwise flow motion of the roll cell and thereby transfers the energy from ${w^\prime}^2$ to $c_{33}$. Such work done by the cross-sectional components of viscoelastic force results in the suppression of ${w^\prime}^2$ at $\Wew=1000$. 

At higher Weissenberg numbers of $\Wew=1500$ and 2000, where the roll cells are modulated to a 2D straight type, the viscoelastic work distribution is more symmetric with respect to the channel centerline and almost zero in the near wall region, unlike the $\Wew=1000$ case. The work is significant instead between the roll cells, where the secondary fluid motion is mainly in the wall-normal direction. 
The in-plane viscoelastic work is positive in the near wall region where the secondary flow converges ahead, away from the wall; and there is a negative force on the other side of the channel, where the secondary flow approaches the wall and diverges. 
As the distribution is antisymmetric, the positive and negative peaks on both sides of the roll cells cancel each other when averaged in the spanwise direction, and the net viscoelastic work $W^{ve}_{22}$ is nearly zero, as shown in \Fref{fig:budget}(b). 

%%%%%%%%% Figure 14 %%%%%%%%%%%%%%%
\begin{figure}
\begin{center}
	\includegraphics[width=0.75\hsize]{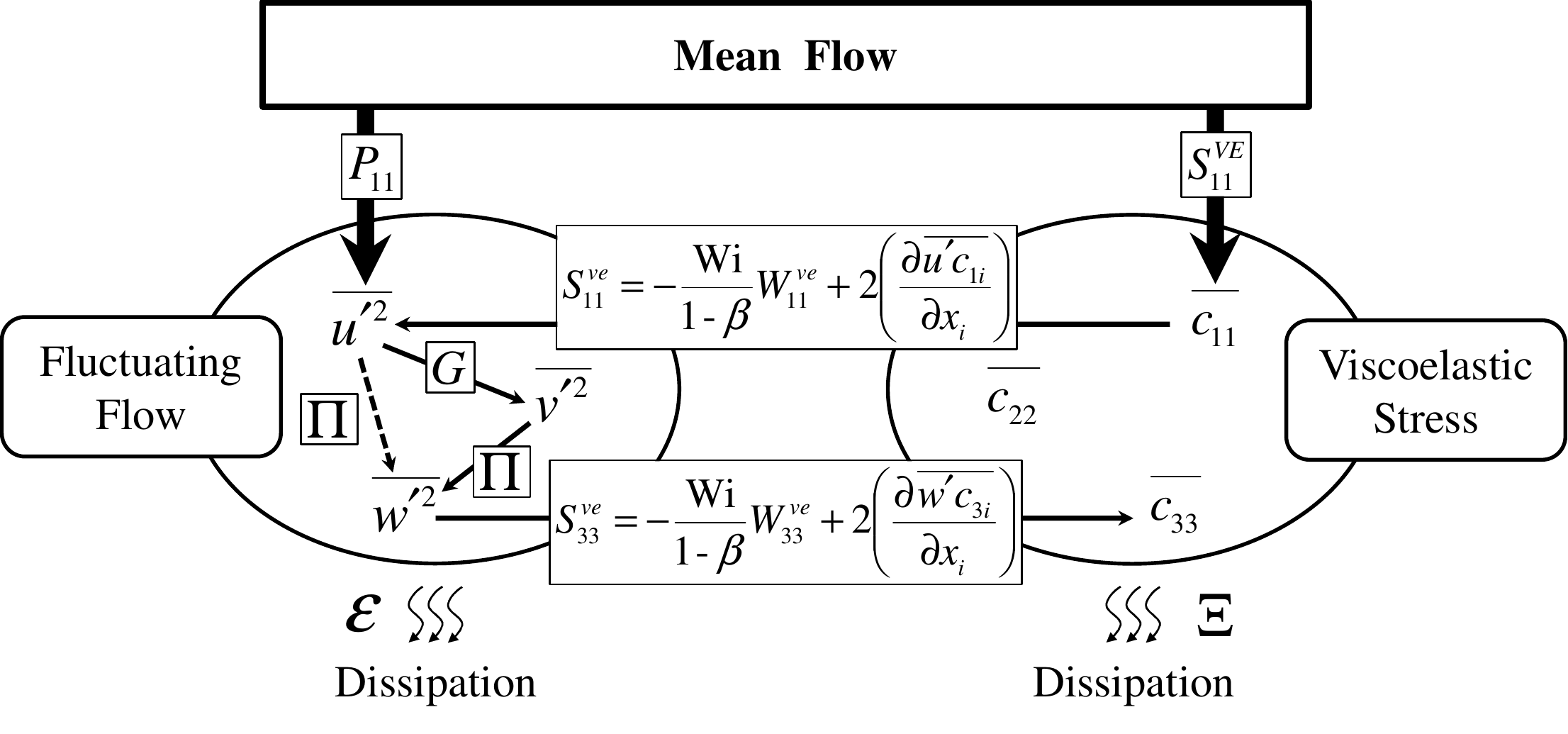}	
	\caption{Schematic of energy exchange between the flow field and viscoelastic stress (conformation tensor) field, indicating the energy cascade from the mean flow to dissipation.}
	\label{fig:schem}
\end{center}
\end{figure}
%%%%%%%%% Figure 14 %%%%%%%%%%%%%%%

Based on the investigation described above, one may draw a schematic view of how the energy is transferred from the mean flow to the fluctuating flow and the conformation tensor fields, and then eventually dissipates, as presented in \Fref{fig:schem}. 
In the flow field, the energy is given from the mean flow to $\overline{{u^\prime}^2}$ at first, then distributed to the other normal components by either the Coriolis-force or the pressure-strain redistribution, and eventually dissipated from each component. 
However, among the mean viscoelastic stresses (or conformation tensor), $\overline{c_{11}}$ first receives energy from the mean flow, but there is no energy redistribution between the normal stress components, unlike the Reynolds stresses, as \Eref{eq:cij} does not have any term that corresponds to the pressure-strain term in the Reynolds-stress transport equation \eref{eq:rss}. 
Although the nonlinear term, including the mobility factor $\alpha \neq 0$, is considered, we may assume its contribution to be negligible compared to the other terms in this study.
In the viscoelastic case, there are two paths for $\overline{{u^\prime}^2}$ to gain energy from the mean flow: one is directly from the mean flow through $P_{11}$, and the other is an indirect way via $\overline{c_{11}}$ through $W^{ve}_{11}$. 
Because of such an additional energy gain via the viscoelastic stress, the high and low speed streaks can maintain their magnitude. 
In contrast, the spanwise secondary flow is weakened as $\Wew$ increases, as the energy is absorbed by the $\overline{c_{33}}$ component, in the case of wavy roll cells. 
In the case of higher $\Wew$ with the 2D roll cells, the energy exchanges between $\overline{{v^\prime}^2}$ and $\overline{c_{22}}$, and between $\overline{{w^\prime}^2}$ and $\overline{c_{33}}$ are insignificant, and only the energy transport from the mean flow to the $\overline{{u^\prime}^2}$, directly and via $\overline{c_{11}}$, occurs. 

In this study, we investigated a viscoelastic laminar wall-bounded flow with streamwise-elongated roll cells and observed that $\overline{{v^\prime}^2}$ and $\overline{{w^\prime}^2}$ decreases with the increasing $\Wew$ whereas the high and low-speed streaks are maintained. 
This is consistent with the experimental observation by \cite{virk75} for the viscoelastic wall turbulence. 
Some explanations have been proposed for the mechanism of such a viscoelastic effect. 
\cite{min03} explained that polymer stretches and thereby absorbs the kinetic energy from the flow when lifted from the near wall region by the vortical fluid motion, and then releases an elastic energy to the flow away from the wall. 
\cite{dubief04} conjectured that the polymer is stretched while it is drawn into near-wall vortical motions and re-injects the energy into the near-wall streaks through polymer work, by which the streaks are enhanced.
We clearly showed that the additive absorbs the kinetic energy from the mean flow through the stretching effect by the mean shear, rather than by being lifted up or drawn into the roll cells, and eventually transfers its elastic energy to the streamwise velocity fluctuation through the additive effect. 
Although a certain force from the secondary flow to the additive is also observed, their net contributions to the energy balance equations are shown to be insignificant or almost zero after averaging. 
Furthermore, it has been shown by \Eref{eq:cij} that there is no inter-component energy transfer between the viscoelastic stresses unlike the Reynolds stresses, which indicates that the energy exchange between the secondary flow and the wall-normal and spanwise viscoelastic-stress components are not related with the streak-enhancing effect of viscoelasticity. 
Hence, the mechanism by which the magnitude of the streaks is maintained can be explained as follows: the additive is significantly stretched near the wall towards the streamwise direction by the shear of the mean flow, absorbing the mean-flow kinetic energy, and transferring the energy to the streamwise velocity fluctuation when advected away from the near wall region by the secondary flow of the roll cells. 
The streaks are, therefore, maintained only by the interaction between the streamwise components of the mean flow, viscoelastic normal stress, and fluctuating flow. 
The secondary flow of the roll cells merely advects the additive. 
 
The viscoelasticity has also been shown to modulate the streamwise-dependent wavy roll cells into streamwise-\emph{independent} 2D roll cells. 
This fact can be interpreted as the streamwise-elongated streak structure being possibly stabilized by the elasticity. 
Similar observations have been reported by \cite{Biancofiore17}, who demonstrated that the viscoelasticity may delay the growth of secondary instabilities of the streaks and eventually the resulting transition to turbulence. 
As proposed by \cite{waleffe97} and \cite{jimenez99}, it is an important part of the autonomous regeneration process of near-wall turbulence that the near-wall streaks become streamwise independent due to the secondary instability. 
The viscoelasticity, therefore, may have an effect in disturbing the self-sustaining process of the wall turbulence by preventing the streaks from being streamwise dependent according to the secondary instability, and yielding a delay of streak breakdown due to turbulent transition.

\begin{table}
\caption{\label{tab:time}Friction Reynolds and Weissenberg numbers, and various timescale ratios.}
\footnotesize\rm
\begin{tabular*}{\textwidth}{@{}*{5}{@{\extracolsep{0pt plus12pt}}l}}
\br
&Newtonian & $\Wew=1000$ & $\Wew=1500$ & $\Wew=2000$\\
\mr
$\Ret = u_\tau \delta / \nu$                  & 15.4 & 13.7  & 14.5 & 14.5 \\
${\rm Wi}_\tau = \lambda u^2_\tau / \nu$      & ---  & 18.8  & 31.6 & 41.7 \\
${\rm Wi}_\tau /\Ret = \lambda u_\tau/\delta$ & ---  & 0.686 & 1.09 & 1.45 \\
$\Wew /\Rew = \lambda U_{\rm w} / \delta$     & ---  & 10.0  & 15.0 & 20.0 \\
\br
\end{tabular*}
\end{table}

Lastly, we discuss the relevance of the drag-reducing phenomenon in the turbulent flow and also examine the timescales relevant to the present flow of modulated roll cells. 
An important mean-flow parameter, the friction Reynolds number $\Ret$ based on the friction velocity $u_\tau = \sqrt{\tau_\mathrm{w}/\rho}$, was calculated for each case, and is presented in Table~\ref{tab:time}. 
In the viscoelastic flows, $\Ret$ decreased more or less with respect to the Newtonian case and, even so, the presently tested range is around 15. 
Such a low $\Ret$ value is basically within the laminar regime of the wall-bounded shear flow, since the present channel-gap width could be too narrow to contain turbulent motions in terms of the wall unit.
However, the roll cell we investigated here may be comparable to the near-wall streamwise vortex. 
Owing to the fact that the latter observed in the turbulent flow would occur at a wall-normal height of $y^+ (\equiv yu_\tau/\nu) \approx 20$ with a radius of $\approx 15(\nu/u_\tau)$, on average, according to \cite{Kim87}. 
In addition, regarding the friction Weissenberg number ${\rm Wi}_\tau$, the values in Table~\ref{tab:time} are on a reasonable order of 10--40. 
The previous DNS on the drag-reducing channel flow demonstrated that moderate drag-reduction rates are obtained on the order of ${\rm Wi}_\tau = 10$ and the maximum rate could be achieved at ${\rm Wi}_\tau = 40$ or more \cite{Housiadas03,tsuka11}. 
This matching of the order of ${\rm Wi}_\tau$ may support our hypothesis that the viscoelasticity-induced modulations occurring in the roll cell would be equivalent to those in the near-wall vortex of the drag-reducing turbulent flow.
The ratio of the friction Reynolds and Weissenberg numbers corresponds to the timescale ratio between the relaxation time and the wall-shear rate ($u_\tau/\delta$).
Given that the rotational speed of roll cell (as well as the near-wall vortex) would be on the order of $u_\tau$, the timescale of $\delta/u_\tau$ can be interpreted as the turnover time. 
Table~\ref{tab:time} shows $\lambda u_\tau/\delta$ close to the unity, while $\lambda U_{\rm w}/\delta \gg 1$ ($U_{\rm w}/\delta$ represents the shear rate of the laminar base flow).
This clarifies that the viscoelastic effect of suppressing the secondary flow and the streamwise dependency, would occur on the roll cell (i.e., the streamwise vortex) having a turnover time comparable to the relaxation time.
Furthermore, it is interesting to note that $\lambda u_\tau/\delta$ is very close to 1 in the case of $\Wew = 1500$, which yields the most stable 2D roll cell; the lower and higher values (than 1) might provide a less effective modulation into 2D structure and induce an instability, respectively, as reported in Section 4. 
One may conjecture that $\lambda u_\tau/\delta$, or ${\rm Wi}_\tau/\Ret$, would be a suitable index for the onset of viscoelastic effect on wavy roll cells in wall-bounded shear flow.

%%%%%%% New Section %%%%%%%%%%%%%%%%%%%%%%%%%%%%%%%%%%%%%%%%%%%%%%%%%%%%%%%%%%%%
\section{Conclusion}
In this study we performed direct numerical simulations of a viscoelastic rotating plane Couette flow at $\Rew=100$ and $\Omega=10$, to investigate the effect of addition of viscoelasticity on the streamwise-dependent wavy roll cells that are observed in Newtonian flows. 
The Weissenberg number was changed up to $\Wew=2000$, which corresponds to the friction Weissenberg number of ${\rm Wi}_\tau \approx 40$ at the given parameter set. The flow remains laminar and accompanied by steady roll cells even for the viscoelastic fluid flow.
It is shown that, as $\Wew$ increases from 1000 up to 2000, the wall-normal and spanwise velocity fluctuations are suppressed while the streamwise fluctuation is maintained, and the wavy roll cells are modulated into streamwise-independent 2D roll cells, while viscoelastic torques indicate counteractions to the wall-normal vorticity and rotating roll cells for the wavy roll cells.
We analyzed the Reynolds-stress transport equations and noted that the additive work terms suppress the spanwise fluid motion in the case of the wavy roll cells by absorbing the kinetic energy of the secondary flow, whereas their effects are insignificant in the case of the streamwise-independent 2D roll cells. 
The mean flow provides a comparable amount of energy to the viscoelastic stress $\overline{c_{11}}$ as it does directly to $\overline{{u^\prime}^2}$, which is eventually further transferred to $\overline{{u^\prime}^2}$, too.  
As there is no inter-component energy exchange between the viscoelastic stress components, the $\overline{c_{22}}$ and $\overline{c_{33}}$ components do not have a significant energy supply, and the additive work terms therefore act as an energy sink in the $\overline{{v^\prime}^2}$ and $\overline{{w^\prime}^2}$ equations, whereby the secondary flow is weakened in the viscoelastic cases. 

Similar effects by elasticity may be also observed in the near-wall structures in the viscoelastic drag-reducing wall turbulence, as the energy transfer from the mean flow in the wall turbulence is also only to the streamwise component of the Reynolds and viscoelastic stress, similarly to the laminar flow with roll cells investigated in the present study. The present friction Weissenberg numbers were consistent with those of the viscoelastic wall turbulence where a significant drag reduction can be achieved. The turnover time of the modulated roll cell was comparable to the relaxation time.

Although the present Reynolds number was considerably lower than the turbulent regime, we demonstrated using DNS on RPCF that the drag reduced flow can be achieved even in organized roll cells with a laminar background and also, that the drag-reducing flow should be associated with an elongated streaky structure with less dependency in the streamwise direction. This may provide a key to understanding the drag-reduction mechanism. The above conclusions have been drawn from very limited cases; however, further DNS studies will be conducted in the near future. As seen in the present highest Weissenberg number flow, the elasticity-induced instability is dominant especially for a low Reynolds number flow. Another DNS study at $\Rew = 25$ reported an onset of unsteadiness in the steady roll cell \cite{nimura17,nimura18}. 
Even at moderate or high Reynolds numbers, the elasticity actually enhances the momentum transport and elasto-inertial turbulence: for instance, \cite{Liu13} reported an enhanced torque in the viscoelastic Taylor--Couette flow. 
It is challenging, but necessary to calculate at higher Reynolds numbers with a wide range of Weissenberg numbers.

\ack
T.K. was supported by a Research Fellowship for Young Scientists \#17J04115 from JSPS (Japan Society for the Promotion of Science). 
This work was partially supported by Grant-in-Aid for Young Scientists (A) \#16H06066 from JSPS.
The present simulations were performed on SX-ACE both at the Cybermedia Centre of Osaka University.

%%%%%%%%%%%%%%%%%%%%%%%%%%%% App. %%%%%%%%%%%%%%%%%%%%%%%%%%%%%%%%
%\section{Appendices}

%%%%%%%%%%%%%%%%%%%%%%%%%%%% Ref. %%%%%%%%%%%%%%%%%%%%%%%%%%%%%%%%
\section*{References}
\bibliographystyle{jphysicsB}

\end{document}